# Anatomy of scientific evolution

Jinhyuk Yun[1], Pan-Jun Kim[2,3*], Hawoong Jeong[1,2,4*]

[1]Department of Physics, Korea Advanced Institute of Science and Technology, Daejeon, Republic of Korea.
[2]Asia Pacific Center for Theoretical Physics, Pohang, Republic of Korea.
[3]Department of Physics, Pohang University of Science and Technology, Pohang, Republic of Korea.
[4]Institute for the BioCentury, Korea Advanced Institute of Science and Technology, Daejeon, Republic of Korea.

Corresponding author:
*Email: pjkim@apctp.org (P-JK); hjeong@kaist.edu (HJ)

## Abstract

The quest for historically impactful science and technology provides invaluable insight into the innovation dynamics of human society, yet many studies are limited to qualitative and small-scale approaches. Here, we investigate scientific evolution through systematic analysis of a massive corpus of digitized English texts between 1800 and 2008. Our analysis reveals great predictability for long-prevailing scientific concepts based on the levels of their prior usage. Interestingly, once a threshold of early adoption rates is passed even slightly, scientific concepts can exhibit sudden leaps in their eventual lifetimes. We developed a mechanistic model to account for such results, indicating that slowly-but-commonly adopted science and technology surprisingly tend to have higher innate strength than fast-and-commonly adopted ones. The model prediction for disciplines other than science was also well verified. Our approach sheds light on unbiased and quantitative analysis of scientific evolution in society, and may provide a useful basis for policy-making.

# Introduction

The history of humankind can be summarized in a series of keywords. From the Palaeolithic Age of stone tools to the Information Age of digital technology, science and technology have played a fundamental role behind keywords such as stone, metal, type printing, internal combustion engine, and Internet. To gain a better understanding of human history, numerous intellectuals have explored innovations in science and technology, e.g., science historians like Thomas Kuhn [1] and futurists like Alvin Toffler [2]. Despite the significant contributions of such endeavours, they are essentially derived from qualitative approaches based on individual's accumulated knowledge, and thus necessitate complementary methodology with a more quantitative and unbiased focus. In another aspect, some scientists have developed statistical measures of scientific impact based on paper citations. Although these measures can quantify the impact of papers [3], authors [4]–[5], and journals [6], they are usually focused on gauging the impact within the research community rather than on society in general. Also, there have been built mathematical models to describe the dynamics of scientific paradigms in the whole society [7], but they instead don't provide much evidence of empirical support. Here, on the basis of empirical data, we attempt systematic and quantitative analysis of scientific evolution in the whole society.

We supposed that an extensive, digitized collection of documents long produced in society might be suitable for such analysis. *Google Books Ngram Corpus* [8]–[9] covers 8,116,746 books, ~6% of all books ever printed from all fields of publication between 1506 and 2008. Specifically, the dataset provides information regarding the number of times a given 1-gram or *n*-gram occurred in the books over time. Here, a 1-gram is a string of characters uninterrupted by a space, e.g., a word or number. An *n*-gram is a sequence of 1-grams, e.g., a phrase with three words is a 3-gram. For simplicity, we focused only on 1-grams from the corpus of English books. We calculated the relative frequency of each 1-gram defined as the number of instances of the 1-gram in a given year divided by the total number of 1-grams in the corpus in that same year. The frequency, therefore, represents how widely a given 1-gram was adopted in the public. In addition, to obtain sufficient statistical power for the analysis, we restricted our study to the years after 1800, when at least 70 million words were available each year. Because the dataset itself doesn't provide information regarding which 1-grams are terminologies for science and technology, we identified them with a reference set of scientific and technological words collected from various sources (7,588 words obtained from a science dictionary, scientific journals, and patents; see Materials and Methods). Multiple inflectional forms with a given word stem, such as singular and plural, were integrated systematically when we counted the 1-gram frequency [10]. Because polysemy and synonymy may affect the frequency profiles [11] and thus mislead our analysis, we tried to minimize the presence of the corresponding words amongst our scientific and technological words (Supplementary Methods and Tables S2–S5 and S7 in File S1). We further assumed the frequency of a given scientific or technological word to be an estimate of how widely the actual scientific concept was adopted in society (Supplementary Methods in File S1). All these procedures allowed us to monitor quantitatively the trajectories of science and technology over the years reflected by the frequency profiles.

One clear advantage of investigating such two-centuries-long data, not available from usual online resources with much shorter periods, is that scientific concepts that became widespread after a lag of enormous time could be identified. For example, "biofuel" and "toxicologist" spent 58 and 166 years, respectively, becoming widely used words. Society's response to a new scientific concept is not always immediate. The origin and significance of such 'late bloomers' are discussed later.

## Results

**Characterization and classification of word trajectories**

To characterize the trajectory for each 1-gram, we introduce three measures – first passage time, lifetime, and peak. First passage time (FPT) is defined as years it took the frequency to exceed a certain cutoff $f_c$ since the onset of the 1-gram, capturing how slowly the 1-gram initially spread into society. Lifetime is defined as years between the first and last time of the frequency over the cutoff $f_c$, indicating how long the 1-gram was commonly adopted by society (see Materials and Methods). Peak is defined as the highest frequency of the 1-gram over the entire time. For FPT and lifetime, we set $f_c = 10^{-7}$, which roughly corresponds to a typical frequency of 1-grams found in published dictionaries (Figure S2 in File S1) [8]. As a result, most 1-grams could be classified into the following three types: type-I includes 1-grams with finite and well-defined lifetimes within the time frame of our data (like "phototube" in Figure 1a; for a detailed definition of 'well-defined lifetimes', see Materials and Methods). Type-II, in contrast, shows a lifetime to a distinctively long extent beyond the time frame, so the exact lifetime cannot yet be determined (like "homeostasis" in Figure 1a). One may claim that the classification of type-I and type-II is merely based on the limited period of observation allowed in our current dataset, and thus incorrectly divides the continuum of 1-gram profiles. Although we cannot entirely exclude that possibility, Figures S4 and S5 in File S1 do show a more fundamental difference between type-I and type-II: the overall frequency distribution of type-II shifts to higher ranges over time, while that of type-I stays almost steady. This intrinsic difference between types-I and –II seems to have a mechanistic ground, as will be discussed later (Figure S15 in File S1). Lastly, type-III, unlike types-I and -II, comprises 1-grams that have not reached any frequency higher than $f_c$, and these words were unlikely to meet in our ordinary life.

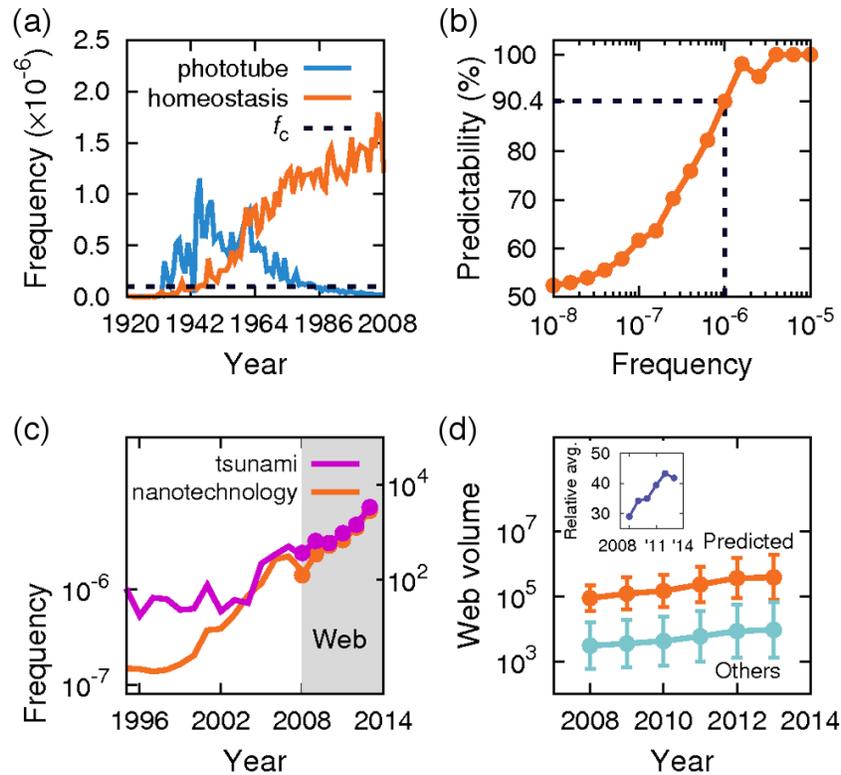

**Figure 1. Classification of scientific words and predictability for long-lasting adoption.** (a) Examples of type-I and type-II scientific words. The vertical axis represents frequency over the years and $f_c$ is a cutoff frequency used for measuring lifetime. (b) Predictability for type-II (precision of prediction), which is defined as the fraction of type-II among scientific words that passed a particular frequency on the horizontal axis before 1920. (c) Examples of scientific words predicted to be future type-II. From 2008, the shaded area is for the outcomes of the Google web search engine: the right vertical axis represents webpage volumes updated annually, normalized by the geometric mean over random scientific words (Supplementary Methods in File S1). Matching of each frequency and normalized webpage volume in 2008 is for visual guidance, not intended to infer a one-to-one correspondence between the two scales. (d) Webpage volumes updated annually since 2008, for all scientific words predicted as future type-II and for other randomly-selected scientific words (Supplementary Methods in File S1). Geometric means are plotted along with error bars from geometric standard deviations. (inset) annual ratio of the geometric mean of the predicted type-II to that of the other random scientific words. In (c) and (d), prediction for type-II was made according to the level of frequency passed between 2000 and 2008.

**Predictability for long-prevailing scientific concepts**

The existence of the above three different types of 1-grams raises an intriguing question: can one predict which science and technology will prove to be type-II (long-term successes) based on levels of prior frequency? By calculating the fraction of type-II among scientific words with each level of frequency exceeded before 1920, we found 90.4% were type-II if a frequency of $10^{-6}$ was passed ($P = 2.3 \times 10^{-20}$; the fraction slightly changes if one considers year $\geq$ 1920 for the frequency being passed. See Supplementary Methods and Figure S6 in File S1). Compared with 61.7% and 52.4% that were type-II for those passing the frequency of $10^{-7}$ and $10^{-8}$, respectively (Figure 1b), 90.4% for $10^{-6}$ is quite noticeable and gives a simple means to predict type-II with high precision based on this frequency of $10^{-6}$. In 1897, for example, "nitroglycerin" passed the frequency of $10^{-6}$, and as currently identified as type-II, has been widely applied to explosives and medicines. As expected, the higher the frequency level crossed by scientific words previously, the more likely they are to be type-II (Figure 1b). Furthermore, for each level of the frequency crossed, scientific words consistently have a larger probability of being type-II than an entire set of 1-grams (including not only scientific words but also the other 1-grams), e.g., the frequency level of $10^{-6}$ involves 90.4% and 35.1% type-II for scientific words and the entire set of 1-grams, respectively. Motivated by such findings, we can anticipate which contemporary scientific concepts will be type-II in the future based on their frequency level between 2000 and 2008. First, "tsunami", a series of huge water waves, rushed to the frequency of $2 \times 10^{-6}$ in 2006. With a 97.1% chance of being type-II ($P = 3.0 \times 10^{-9}$), we predict that "tsunami" will hit our society for a long time (Figure 1c). Although the fate of the word "tsunami" may be somehow affected by the actual incidence of tsunamis in the future, we notice the tsunamis' socio-economic implications, not just limited by specific tsunami events. Also, "bioethics" crossed the frequency of $1.5 \times 10^{-6}$ in 2007 and will continue to receive the spotlight according to our expectation [12]. We observe the rapid rise of "nanotechnology" (Figure 1c) and practical outcomes of biotechnology, such as "biomarker" and "biosensor". Although not explicit, aging seems to be an important consensus of several rising words such as "osteoarthritis" (degenerative arthritis) and "nephropathy" (kidney disease) [13]–[14]. Cancer and neurological diseases, partially relevant to aging as well, will also live with us for a long time, according to our prediction (see Tables S2–S5 in File S1 for the detailed list).

Note that our prediction is based on the 1-gram dataset available up to 2008. To test how accurate the prediction results can be with a separate up-to-date dataset, we obtained the Internet webpage volumes (as a proxy for word usage) updated annually for scientific words between 2008 and 2013 (e.g., Figure 1c for "tsunami" and "nanotechnology"; see Materials and Methods). Indeed, overall webpage volumes of scientific words predicted as future type-II consistently exceed those of other random scientific words by an order of magnitude in the years between 2008 and 2013 (Figure 1d). On average, the ratio of such webpage volumes between the type-II-predicted words and the random counterparts even increases by 44.1% from 29.0 to 41.8 in the same period, indicating the divergence between their growth patterns (Figure 1d inset). We therefore conclude that our prediction works well beyond the time frame of our 1-gram data.

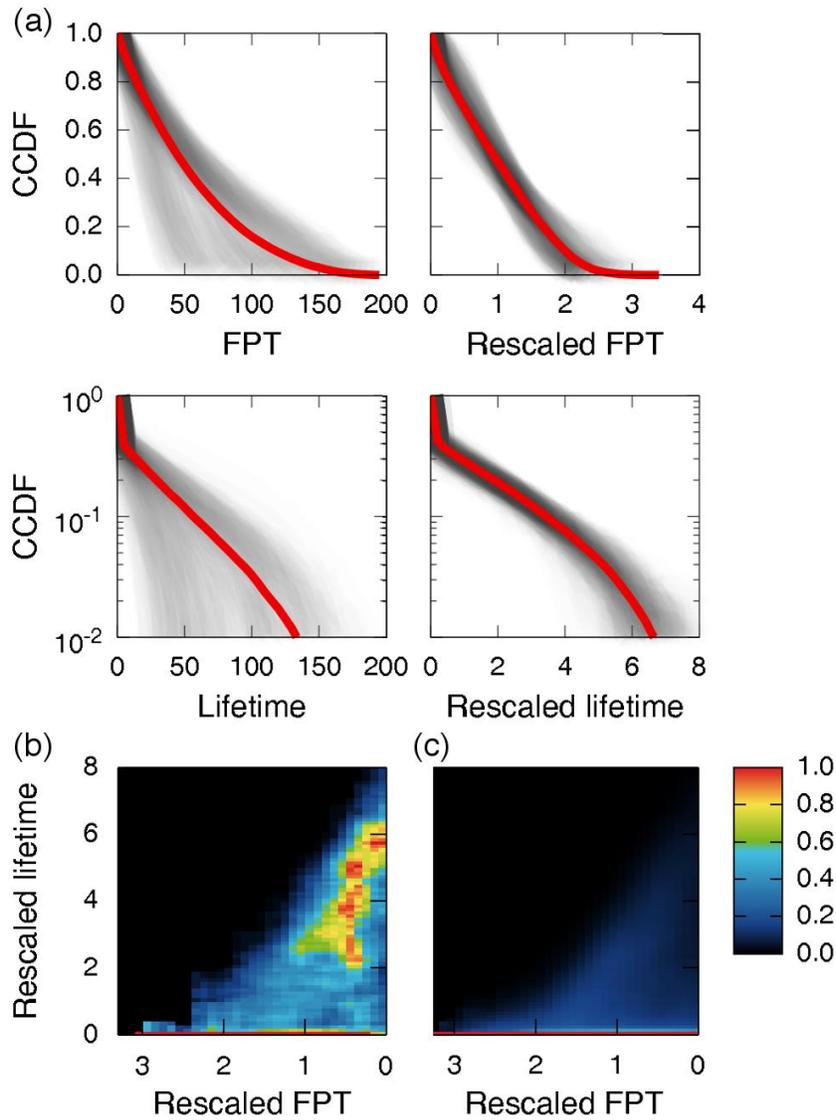

**Figure 2. Characteristics of first passage time (FPT) and lifetime.** (a) Complementary cumulative distribution functions (CCDFs) of FPT, lifetime, and their rescaled values for type-I 1-grams. Shaded areas correspond to the CCDFs with each for 1-grams from the same year of birth. Each red line denotes the CCDF for all 1-grams aggregated from different years of birth. (b, c) Density plot between rescaled FPT and lifetime in the type-I case, for scientific words (b) or for an entire set of 1-grams (c). We hereafter call the rescaled FPT and lifetime from the data simply FPT and lifetime. Each spot is coloured according to the density of 1-grams at the corresponding FPT and lifetime. Specifically, for each value of FPT, we normalized every density relative to the maximum across lifetime, and according to this adjusted density, coloured the spot following the scale bar on the rightmost side (see Supplementary Methods in File S1).

**Tipping point of scientific evolution**

In order to proceed to in-depth analysis of scientific evolution, we stress the fact that the overall FPT and lifetime of 1-grams were getting shorter over the past years (Figures S7 and S8 in File S1), indicating the acceleration of cultural turnover over time as reported in the original study of *Google Books Ngram Corpus* [8]. This global effect of accelerating 'time' itself makes it unfair to directly compare FPTs or lifetimes many years apart. To compensate for such accelerating effect, we propose the rescaled measures of FPT and lifetime, which now lead to very similar patterns across years (Materials and Methods; see Figure 2a and Figure S7 in File S1). Therefore, the rescaled measures are almost free from the temporal acceleration effect, making it possible to recruit numerous 1-grams from different years into the same place for analysis. For FPT and lifetime from the data, we hereafter use their rescaled values unless specified.

A logical step forward is to search for any possible interplay between FPT and lifetime in scientific evolution, regarding the long-term effect of initial adoption rates inversely captured by FPT. One can suppose that lifetime varies gradually as a function of FPT through the progressive long-term effect of FPT. Unexpectedly, we discover that type-I scientific words undergo a sudden transition from unimodality to bimodality in their lifetime at a particular value of the FPT. The bimodality at this transition (FPT~1.2) is characterized by two prominent lifetimes of ~2.0 and < 0.1, while the unimodality is characterized by < 0.1 (see Figure 2b). In other words, once initial adoption rates are even slightly higher than a particular value, type-I scientific words may possibly exhibit sudden leaps in their eventual lifetimes ($P = 4.3 \times 10^{-47}$). However, an entire set of type-I 1-grams, which includes not only scientific words but also the other 1-grams comprehensively, doesn't show such behavior (Figure 2c). Besides the case of FPT, an increase in peak leads to a similar transition of lifetime for scientific words, but does so for an entire set of 1-grams barely at much larger peak, 11.3 times as large as scientific words (Figure S10 in File S1). Taken together, the results demonstrate that the temporal evolution of science and technology is subject to an *abrupt* transition at a threshold or 'tipping point'. The possible mechanism behind the transition will be addressed below, through our mathematical modeling.

**Mechanistic model of scientific evolution**

To understand the underlying dynamics of the observed patterns, we start by identifying three key factors that drive the adoption of science and technology. First, there is preferential adoption. People are more likely to adopt already widespread, popular items than to adopt less popular ones because of a variety of psychological, sociological, and economical reasons [15]–[16], possibly resulting in the rich-get-richer phenomena of innovation spread. Second, the adoption of innovations may also be affected by homophily [17], according to which innovations are more likely to spread among people with similar interests or similar professions. Therefore, newly-introduced science and technology are likely to be shared easily within the scientific community itself rather than between the scientific community and the other communities. Third, every innovative item has its own intrinsic value or fitness, which confers an inherent difference to the item's adoption rate from that of another [18]–[19]. Here we bypass the need to dissect fitness into its detailed constituents, and rather view it as a collective quantity accounting for people's response to an item.

By incorporating the above three factors, we created a mechanistic model of innovation spread. The model comprises *N* agents where the individual agents represent various forms of social units. Agents can invent and adopt items, and the items are transmitted stochastically [20] from agent to agent. Every item is classified into either the scientific category or other, and every agent has the capacity to adopt a total of *L* different items. We further assume that the number of agents, who adopt a particular item, is correlated with that item's frequency in the 1-gram dataset. In other words, the word frequency in the 1-gram dataset is modeled by the item's prevalence among the agents. In the model, the items are adopted through a pre-assigned network between agents as follows. One agent *i* accepts an item $q_j$ of its nearest neighbour agent *j* in the network provided that agent *i* has never adopted the item $q_j$ before [7]. The item $q_j$ subsequently replaces the item $q_i$ of the closest category in the agent *i* with the following probability:

$$P(q_i, q_j, i, j) = f(\lambda_{q_j} - \lambda_{q_i}) \times p(q_j, i) \times p(q_j, j), \tag{1}$$

where $\lambda_{q_{i(j)}}$ is the item $q_{i(j)}$'s fitness, $f(\lambda_{q_j} - \lambda_{q_i})$ is an increasing function of the fitness difference $\lambda_{q_j} - \lambda_{q_i}$, and $p(q_j, i) \times p(q_j, j)$ reflects the effect of preferential adoption and homophily. Specifically, $p(q_j, i)$ takes the following functional form:

$$p(q_j, i) = \sqrt{\frac{\sum_r \delta(q_j, r) w(|\varepsilon_i - \varepsilon_r|)}{\sum_r w(|\varepsilon_i - \varepsilon_r|)}}, \qquad (2)$$

where $\delta(q_j, r)=1$ if agent $r$ has the item $q_j$, otherwise, $\delta(q_j, r)=0$, and $\varepsilon_{i(r)} = 1$ if agent $i$ ($r$) belongs to the scientific community, otherwise, $\varepsilon_{i(r)} = 0$. $w(|\varepsilon_i - \varepsilon_r|)$ is a decreasing function of $|\varepsilon_i - \varepsilon_r|$. The frequency of an item is defined as the ratio of the item's copy number to the total counts of items ($= N \times L$) in the system. For more details of the model, see Materials and Methods.

For both scientific and other items, the mechanistic model captures essential features of empirical relationship between FPT and lifetime in the type-I case (Figure 3a and b; compare them with Figure 2b and c) as well as manifests distinctively long lifetime for type-II (Supplementary Methods and Figures S12–S15 in File S1). Specifically, preferential adoption and homophily are crucial to demonstrate the splits of lifetime into different groups: a separation of type-I and type-II, and an abrupt transition in type-I scientific items. Without preferential adoption and homophily in the model, these splits are hard to observe (Supplementary Methods in File S1). Fitness is also important in our model. Without fitness, the model fails to produce the diagonal structure that lies in the ranges of rescaled FPT ≤1.2 and rescaled lifetime ≥2.0 in Figure 2b (Supplementary Methods in File S1). Therefore, three key components in the model – preferential adoption, homophily, and fitness – are important toward explaining the observed patterns in scientific evolution. Interestingly, according to the model, type-I and type-II scientific items are adopted longer in the opposite places, type-I in the scientific community and type-II in the outer society (Figure S15 in File S1).

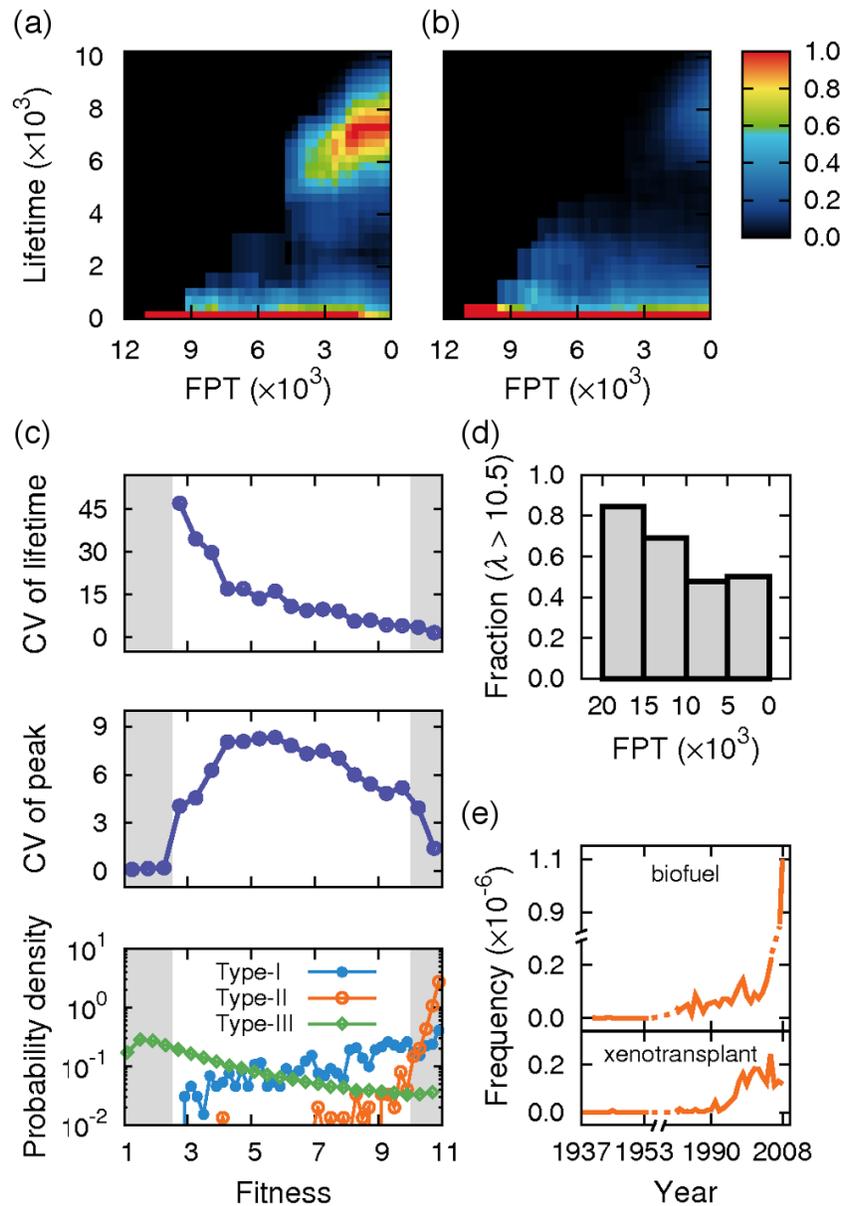

**Figure 3. Model simulations and late bloomers.** (a, b) Density plot between FPT and lifetime in the type-I case, for scientific items (a) or else (b) from the model simulation. Coloured in the same way as Figure 2b and c. (c) Uncertainty in the long-term fate of science and technology. For each value of fitness, plotted are the coefficient of variation (CV) of lifetime (top), CV of peak (middle), and the probability densities of types-I, -II, and -III (bottom). CVs of lifetime and peak were obtained from all three types by defining the lifetime of type-III as zero. The shaded area on the top left side includes only type-III, clearly having a uniform lifetime (of zero) in spite of ill-defined CV. Therefore, in the top and middle panels, intermediate fitness shows larger uncertainty of lifetime and peak than low and high fitness of the shaded areas. (d) For each range of FPT,

the fraction of high fitness (fitness $\lambda > 10.5$) among scientific items with well-defined finite FPT (i.e., types-I and -II). (e) Empirical examples of late-bloomer scientific words. Both "biofuel" and "xenotransplant" belong to type-II, with ~60 years passed to reach the frequency of $10^{-7}$ since their birth. The model simulations in (a–d) were performed under the parameters described in Materials and Methods.

**Determinism versus contingency, and late bloomers**

The accomplishments of our model encourage us to address mechanistic issues in science history otherwise difficult to do. The history of science and technology can be seen from two different viewpoints, determinism versus contingency [21]. Relating to these viewpoints, to what extent does the fitness considered in the model 'determine' the success of individual science and technology? Both lifetime and peak, indicators of long-term success of scientific items, increase, on average, as functions of fitness (Figure S16 in File S1). However, the average trend itself doesn't indicate how deterministic it is, and the variability of individual items out of such average trend requires examination. We found that, against the averages at given fitness, lifetime and peak are the most variable at the intermediate level of fitness, while they are less variable, more deterministic at high- and low-level fitness (Figure 3c). Consistently, we observe that type-II (-III) scientific items have a distribution much biased to high-level (low-level) fitness, making this fitness regime less variable (Figure 3c).

In addition to lifetime and peak, FPT draws our attention to its relationship with fitness. Because type-III never attains a frequency higher than the cutoff $f_c$, its FPT is ill-defined and can be regarded as infinite. Type-III, namely having infinite FPT, occupies a larger fraction as fitness gets lower (Figure 3c). This fact, as well as common intuition, suggests an inverse relation between FPT and fitness for types-I and -II having well-defined finite FPT. Contrary to this expectation, we discover that types-I and -II with long FPT surprisingly tend to have higher fitness than those with short FPT (Figure 3d and Figure S19 in File S1). Indeed, in Figure 3d, 72.7% of long FPT $>10000$ are associated with high fitness $>10.5$, while only 49.6% of shorter FPT are associated with that high fitness ($P = 5.7 \times 10^{-8}$). What makes slowly-adopted, long-FPT science and technology have high fitness? The reason, briefly, lies in the fact that high-fitness helps the science resist even long hard times of frequency $< f_c$, yielding long FPT as well as short FPT. In contrast, low-fitness science is difficult to sustain unless it initially spreads rapidly, either acquiring short FPT or falling to type-III (Figure S20 in File S1); 'late bloomers' are permitted by high fitness rather than by low fitness. Besides the model results, *Google Books Ngram Corpus* contains a number of actual late bloomers in science and technology. For example, "biofuel" crossed the frequency of $10^{-7}$ in 2004, 58 years after its birth, involving renewable energy and environmental issues (Figure 3e) [22]. "isoflavone", a compound in soybean, required 70 years to reach the same frequency, and is receiving attention for its anti-cancer effects [23]. Also, "toxicologist" had to wait even 166 years until it met a frequency of $10^{-7}$ in 1975. In medicine, "xenotransplant", animal tissue or organ transplant in a human patient, was initially believed to work hardly due to immunologic barriers [24], but eventually succeeded in passing a frequency of $10^{-7}$ in 1997, 61 years after the birth (Figure 3e). Table S7 in File S1 presents a more comprehensive list of late bloomers observed in scientific evolution.

**Verification of the model prediction for other disciplines**

Although our model was primarily intended to account for the observed patterns in scientific evolution, we notice that three key components of the model − preferential adoption, homophily, and fitness − can also be valid for the evolution of other professional fields driven by innovation diffusion between the specialized community and the public. For any fields with these three key components, our model suggests that the relationship between FPT and lifetime for type-I is similar to that shown in Figure 2b. In this regard, food and art may be good candidate fields to test the prediction. The words in food and art [25]–[27] indeed follow the predicted patterns in their FPT and lifetime (Figure 4a and b; $P = 3.1 \times 10^{-9}$ for food and 0.018 for art). The results are robust to the exclusion of words overlapping with those analysed for scientific evolution (Figure S21 in File S1), supporting the empirical validity of the key components in our model.

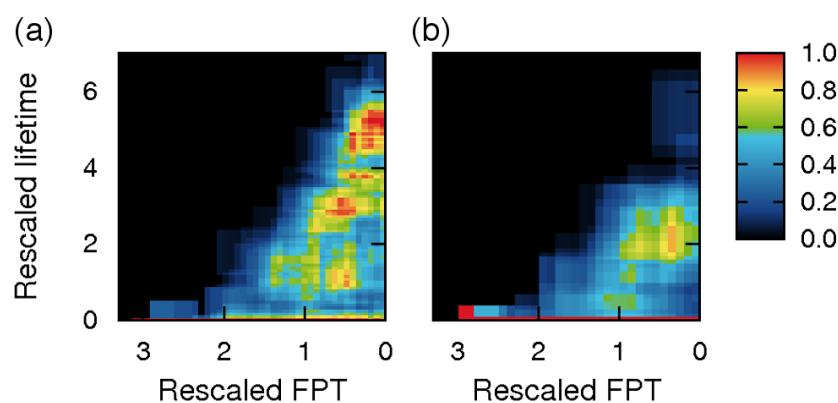

**Figure 4. Analysis of other fields: food and art.** (a) Data from food and nutrition [25]. (b) Data from art and music [26]–[27]. In (a) and (b), density plot between rescaled FPT and lifetime for type-I, coloured in the same way as Figure 2b.

## Discussion

In this study, we explored the evolution of science and technology through a massive corpus of digitized English texts over the past two centuries, highlighting the whole society's influence beyond that of the specialized community (Figure S15 and Tables S2–S5 and S7 in File S1). Scientific evolution is not solely driven by the isolated action of scientists but by the collaboration between scientists and society. We suggest that in-depth analysis into a causal or feedback relation between scientific research and word usage in society may be warranted to enhance the impact of our approach. Also, extending our analysis to $n$-grams with $n>1$ and refining the presented model are left for further study.

Our approach has significant implications for policy-making, especially when complemented by other sophisticated methodologies [28]. Governments and institutions often agonize over the optimal allocation of research resources and incentives to promote good research outcomes [29]. While evaluations for such investments are conventionally based on scholarly outcomes, e.g., the number of publications, patents, and citations, and the reputations from colleagues [3]–[6], [30], the comprehensive impacts of whole research outcomes outside the professional community have recently begun to be appreciated [31]. Beyond the contents of the printed books that we harnessed in this study, modern information society offers a myriad of online resources to check people's response to particular science and technology, such as comments in social media, website hits, media exposure, and blog postings [32]–[34]. In addition, the existence of late bloomers necessitates active consideration of old but recently growing technology for future investment. Going one step forward, if data-driven analysis accompanied by mathematical modelling is judiciously combined with the context-specific perspectives of traditional approaches, the resulting synergy will facilitate an innovative transformation of methodologies in social sciences, humanities, and policy-making.

## Materials and Methods

**Preprocessing of *Google Books Ngram Corpus***

We use the data of *n*-gram counts in the English section of the *Google Books Ngram Corpus* Version 2 [9]. An *n*-gram is a set of *n* successive 1-grams, in which 1-gram is a string of characters uninterrupted by a space. Here, we focus only on 1-grams for simplicity.

The frequency of a 1-gram is defined as the number of occurrences of the 1-gram in a given year divided by the total number of 1-grams in that year. To consider various inflectional forms of words when computing the frequency, we systematically integrated 1-grams by *Porter Stemming Algorithm* [10]. Moreover, we restricted our analysis to the years after 1800 because the quantity of data before 1800 is insufficient to analyse. *Google Books Ngram Corpus* occasionally assigned 1899 or 1905 to books with unknown publication dates [8]. Therefore, for any 1-gram that appeared in the years 1899 and 1905, the frequency was substituted by the average frequency of ±1 years around those years. We also filtered out some 1-grams subjected to possible errors from the optical character recognition (OCR) processes (see Supplementary Methods in File S1).

**Identification of scientific and technological words**

To identify 1-grams belonging to the vocabulary of science and technology, we built a list of science and technology words (as a reference set) from an online science dictionary *"AccessScience"* [35]. However, contemporary dictionary may be rather biased to words that are commonly used today. In order to reduce such bias, we further collected words from various sources covering a wide range of time, including patent grant texts in the *United States Patent and Trademark Office* [36] and titles of articles in scientific journals (Table S1 in File S1). We selected only nouns among those words (Supplementary Methods in File S1). Because frequent usage within the scientific sources was usually for scientific and technological words, we inspected randomly sampled words ($\geq$ 10% coverage for journals, $\geq$ 1% coverage for patents) along the descending order of usage level within each source, and selected all words of the usage level having at least an 80% chance of being scientific and technological words which are not used in too broad a context. If this cutoff covered all words occurring in that source, then we excluded words that were used only once in the source.

**Characterization of $f_c$, FPT, lifetime, peak, and different types of 1-grams**

We use the cutoff frequency $f_c$ as the threshold above which a 1-gram can be roughly considered to be common in society. As the quantification of first passage time (FPT) and lifetime depends on $f_c$, an appropriate choice of $f_c$ is important, and we choose $f_c = 10^{-7}$ which roughly corresponds to a typical frequency of 1-grams in published dictionaries [8]. However, our main results do not qualitatively change as long as $10^{-8} \leq f_c \leq 2\times10^{-7}$. For a given 1-gram, first passage time (FPT) is defined as years it took the frequency to cross $f_c$ since the birth of the 1-gram, lifetime is defined as years between the first and last year of the frequency above $f_c$, and peak is defined as the highest frequency of the 1-gram over time. Specifically, we define lifetimes only for 1-grams that never exceed the frequency $f_c$ for at least 10 years until the end time of the data, because they are rarely expected to bounce back (Figure S1 in File S1). If the frequency crosses and falls into $f_c$ more than once, we consider the latest event of the falling into $f_c$ as the end of the lifetime.

Most 1-grams can be classified into the following three types. Type-I 1-grams have well-defined finite lifetimes as described above. Type-II shows a lifetime to a distinctively long extent beyond the time frame of the data, so the exact lifetime cannot presently be defined. Finally, type-III includes 1-grams that never had a frequency higher than $f_c$.

**Internet webpage volume**

Because the frequency data from *Google Books Ngram Corpus* is limited until the year 2008, we used the outcomes of the Google web search engine for an alternative up-to-date dataset to test the validity of our type-II prediction results. We collected the Internet webpage volumes updated annually, between the years 2008 and 2013, for the words of our search queries (see Supplementary Methods in File S1 for more details). Because Google itself provides search results based on a stemming algorithm, we searched the singular forms of the words instead of their stems. This work was done manually, regarding the policy of Google, which does not permit automatic search queries by web robots.

**Rescaled measures of FPT and lifetime**

We found that the overall FPT and lifetime of 1-grams were getting shorter over the past years (Figures S7 and S8 in File S1). To 'normalize' FPT and lifetime from such accelerating effect, we employed their rescaled measures, $\tau^*$ for FPT and $T^*$ for lifetime:

$$\tau^* = \frac{\tau}{\langle \tau \rangle_y}, \qquad T^* = \frac{T}{\langle T \rangle_y}, \tag{3}$$

where $\tau$ and $T$ are FPT and lifetime of a given 1-gram, respectively, and $\langle \tau \rangle_y$ and $\langle T \rangle_y$ are the averages of FPT and lifetime over all 1-grams in type-I with the same year of birth. For FPT and lifetime from the data, we used their rescaled values unless specified."

**Model construction and simulation**

To account for our data analysis results, we built a mechanistic model incorporating preferential adoption, homophily, and fitness, which are described in the main text. The model is based on information spread among $N$ agents. Each agent represents an individual or a social cohort, which invents and adopts items. Every agent has the capacity to accommodate a total of $L$ different items. The items are transmitted from agent to agent, and we assume that the adopted ranges of such items are projected into the actual usage levels of the corresponding words in our 1-gram dataset [8].

Every agent is assigned $\varepsilon$, which characterizes the level of involvement in specialized areas. In general, $\varepsilon$ can be a vector with real-number components, and here, we only consider the case of scalar binary numbers: $\varepsilon = 1$ if the agent belongs to the scientific community, otherwise, $\varepsilon = 0$. At the beginning of the simulation, $\varepsilon$ is assigned to each agent with a chance of $\rho$ for $\varepsilon = 1$. Once $\varepsilon$ has been assigned to an agent, either $\varepsilon = 1$ or $0$, it never changes during the simulation. At every time step, a new item is invented by a randomly-selected agent $m$ with probability $\alpha$, and this item belongs to the category following the inventor's $\varepsilon$ (i.e., $\varepsilon_m$). The item is also assigned fitness $\lambda$, a positive real number chosen from a given probability distribution [a power-law $\sim (\lambda/\lambda_{min})^{-\gamma}$ for Figure 3a–d; we also considered the Gaussian distribution as described in Supplementary Methods in File S1]. This new item now replaces one of agent $m$'s old items in the closest category. Next, we randomly select a pair of agents $i$ and $j$, among the nearest neighbours in a pre-assigned network structure for innovation spread. Agent $i$ accepts agent $j$'s item $q_j$ if agent $i$ has never adopted the item $q_j$ before [7], and the item $q_j$ subsequently replaces the item $q_i$ of the closest category in the agent $i$ with the probability $P(q_i, q_j, i, j)$ in equation 1. In the case of Figure 3a–d, the network structure between agents was made according to the Erdős–Rényi model [37], specifically, a $G(N, p_{ER})$ model, where each agent was randomly connected to another with probability $p_{ER}$ [38]. We also considered other network structures with a power-law degree distribution [39], but our main results did not change much against the different network structures. In equation 1 for $P(q_i, q_j, i, j)$, $f(\lambda_{q_j} - \lambda_{q_i})$ is an increasing function of the fitness difference $\lambda_{q_j} - \lambda_{q_i}$, and $p(q_j, i) \times p(q_j, j)$ represents the effect of preferential adoption and homophily. For the case of Figure 3a–d, we employed

$$f(\lambda_{q_j} - \lambda_{q_i}) = \frac{1}{2} + \left| \frac{\lambda_{q_j} - \lambda_{q_i}}{10} \right|^\beta \mathrm{sgn}(\lambda_{q_j} - \lambda_{q_i}). \qquad (4)$$

In equation 2 for $p(q_j, i)$, a square root appears because it makes $p(q_j, i) \times p(q_j, j)$ linearly proportional to the population having the item $q_j$ in the case that $\varepsilon$'s are identical for all agents.

$w(|\varepsilon_i - \varepsilon_j|)$ in equation 2 represents the effect of homophily, and is a decreasing function of $|\varepsilon_i - \varepsilon_r|$. Here, we employed $w(|\varepsilon_i - \varepsilon_j|) = \exp[-(\varepsilon_i - \varepsilon_j)^2]$.

At every $N \times L$ steps of simulation, the frequencies of all items in the system were recorded. The frequency of an item is defined as the ratio of the item's copy number to the total counts of items ($= N \times L$) in the system. Here, we use such $N \times L$ steps as the arbitrary unit of time to measure the FPT and lifetime of items. In Figure 3a–d, we present the simulation results with parameters $\gamma = 2.0$, $\beta = 1/4$, $N = 4096$, $L = 10$, $p_{ER} = 0.1024$, $\rho = 0.2$, $\alpha = 0.0001$, and $f_c = 0.00025$. We identified a range of parameters in which our main results remained robust. See Supplementary Methods in File S1 for full details of our model and parameters.

**Statistical significance test**

To test the statistical significance of our results in Figure 1b, we performed a two-sided $Z$-test under the null hypothesis that there is no association between the frequency level and the probability of type-II. For Figure 3d, we conducted a similar analysis under the null hypothesis that there is no association between FPT and the fraction of scientific items with fitness $> 10.5$.

For Figure 2b, we tested the statistical significance of a sudden leap into ~2.0 in lifetime at FPT ~ 1.2. We constructed a $2 \times 2$ contingency table displaying the numbers of the words at FPT $\geq 1.2$ and $< 1.2$, and lifetime $\geq 2.0$ and $< 2.0$. Then, we computed $P$-values based on the Pearson's Chi-squared test. We also conducted similar analyses for Figure 3a and Figure 4.

## Acknowledgments

We thank Buhm Soon Park, KiHong Chung, Yongjoo Baek, Soo-yeon Hwang and Jaeyun Sung for useful discussions. This work was supported by the National Research Foundation of Korea through Grant No. 2011-0028908 (J.Y. and H.J.) and NRF-2012R1A1A2008925 (P.-J.K).

# Anatomy of scientific evolution

Jinhyuk Yun, Pan-Jun Kim, Hawoong Jeong

# Supporting Information

# Table of Contents

# Supplementary Figures

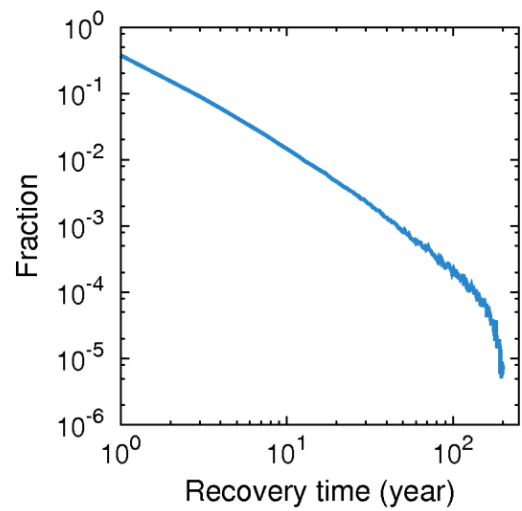

**Figure S1.** Probability distribution of recovery time for 1-gram frequency that fell below $10^{-7}$ and recovered later. The fraction of recovery time monotonically decreases as the recovery time increases. As a result, most cases (82%) have a recovery time < 10 years.

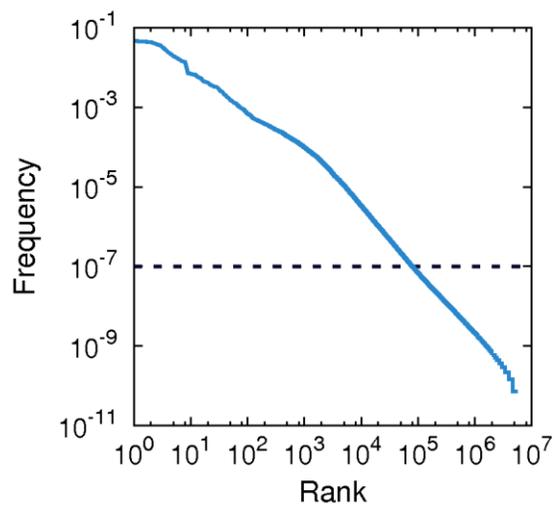

**Figure S2.** Rank-frequency plot for 1-gram stems in the year 2000. Above the frequency of $10^{-7}$ (dashed line), there are 79,691 stems, which are of the same order as the number of stems in an average published dictionary.

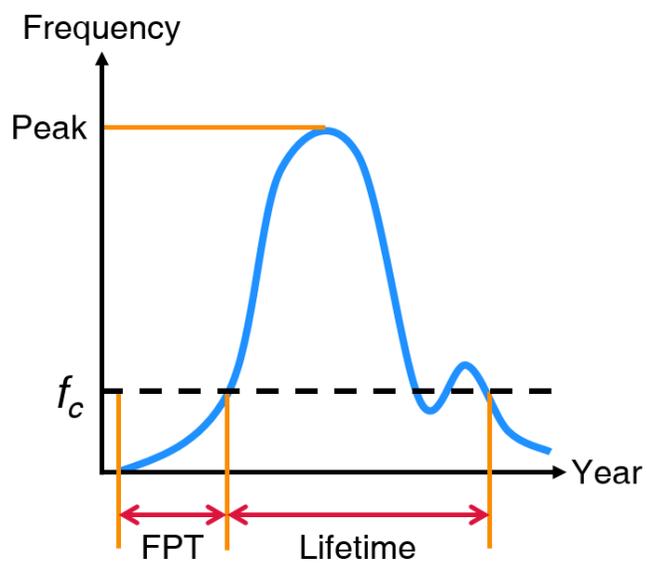

**Figure S3.** Definitions of first passage time (FPT), lifetime, and peak.

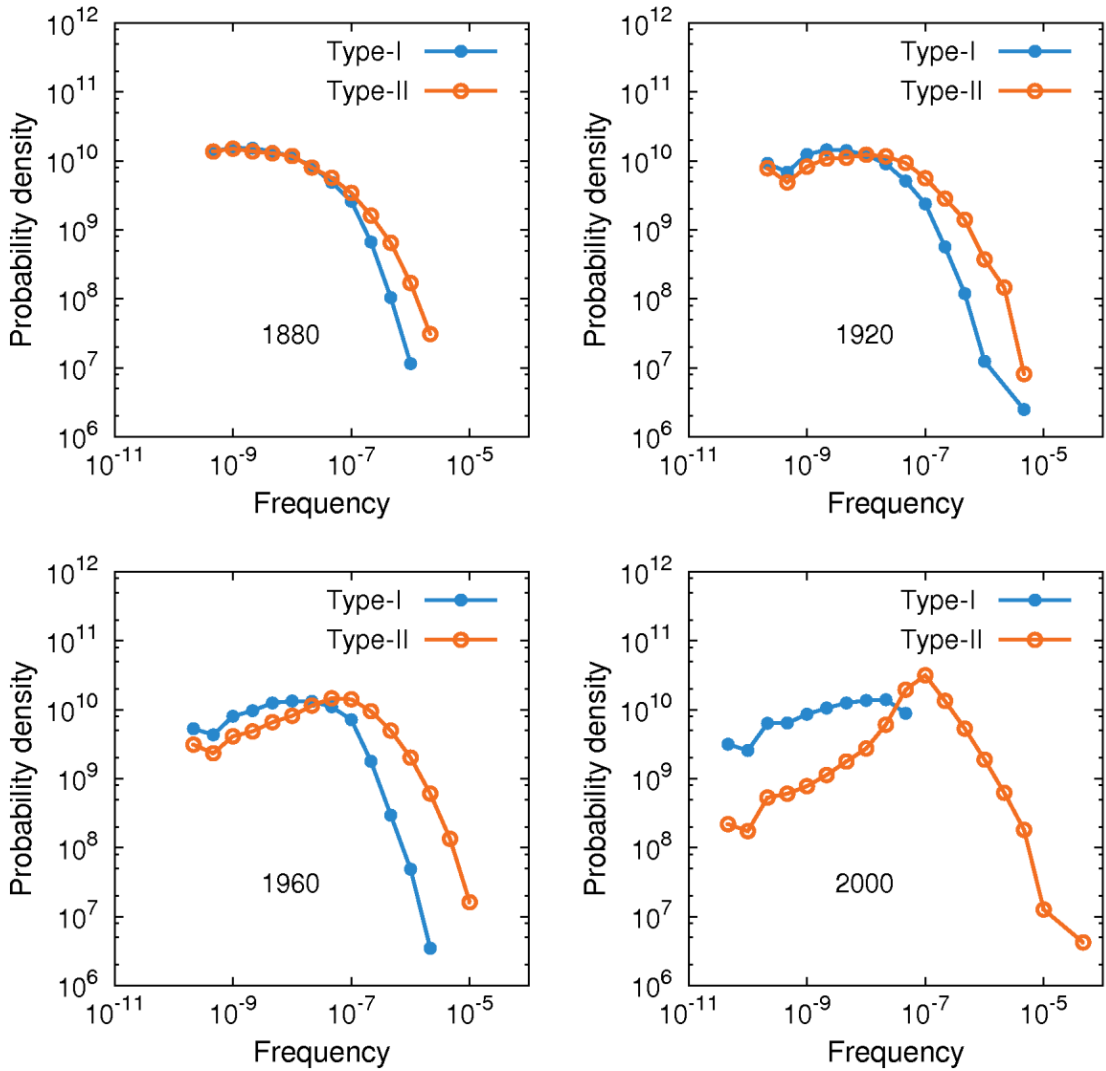

**Figure S4.** Frequency distribution for types-I and -II in each year. In 1880, the probability density functions (PDFs) of types-I and -II almost overlap. As time elapses, the PDF for type-II shifts to higher frequency ranges, while that for type-I stays in almost the same frequency range.

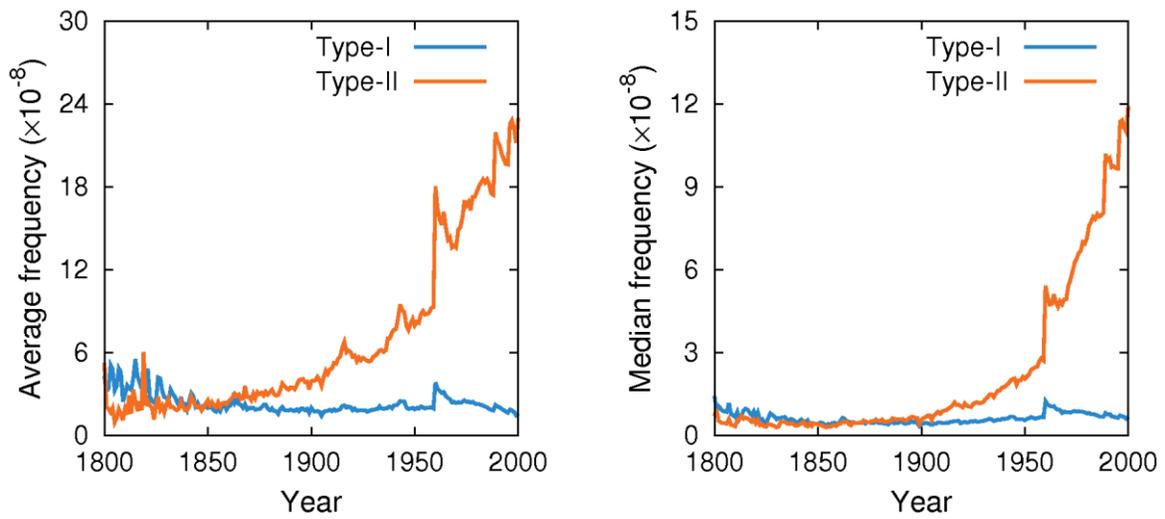

**Figure S5.** Average and median frequencies of 1-grams in types-I and -II over the years. Both the average and median frequencies for type-II increase over time, whereas those for type-I barely change.

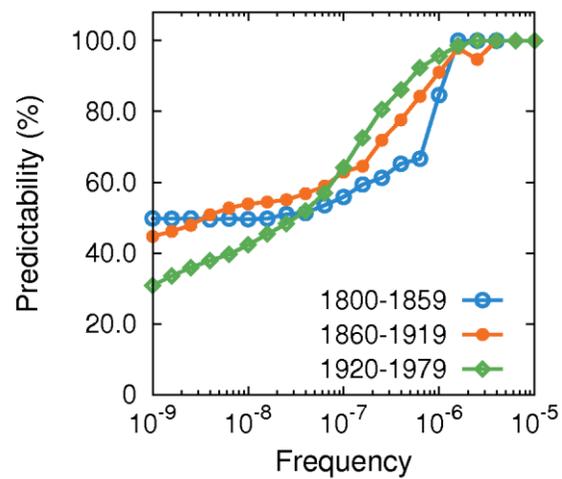

**Figure S6.** Probability that a scientific word turns out to be type-II as of 2008 if it first passed a particular level of frequency on the horizontal axis in a given range of the past years. The higher frequency a scientific word exceeds, the more likely it is of type-II.

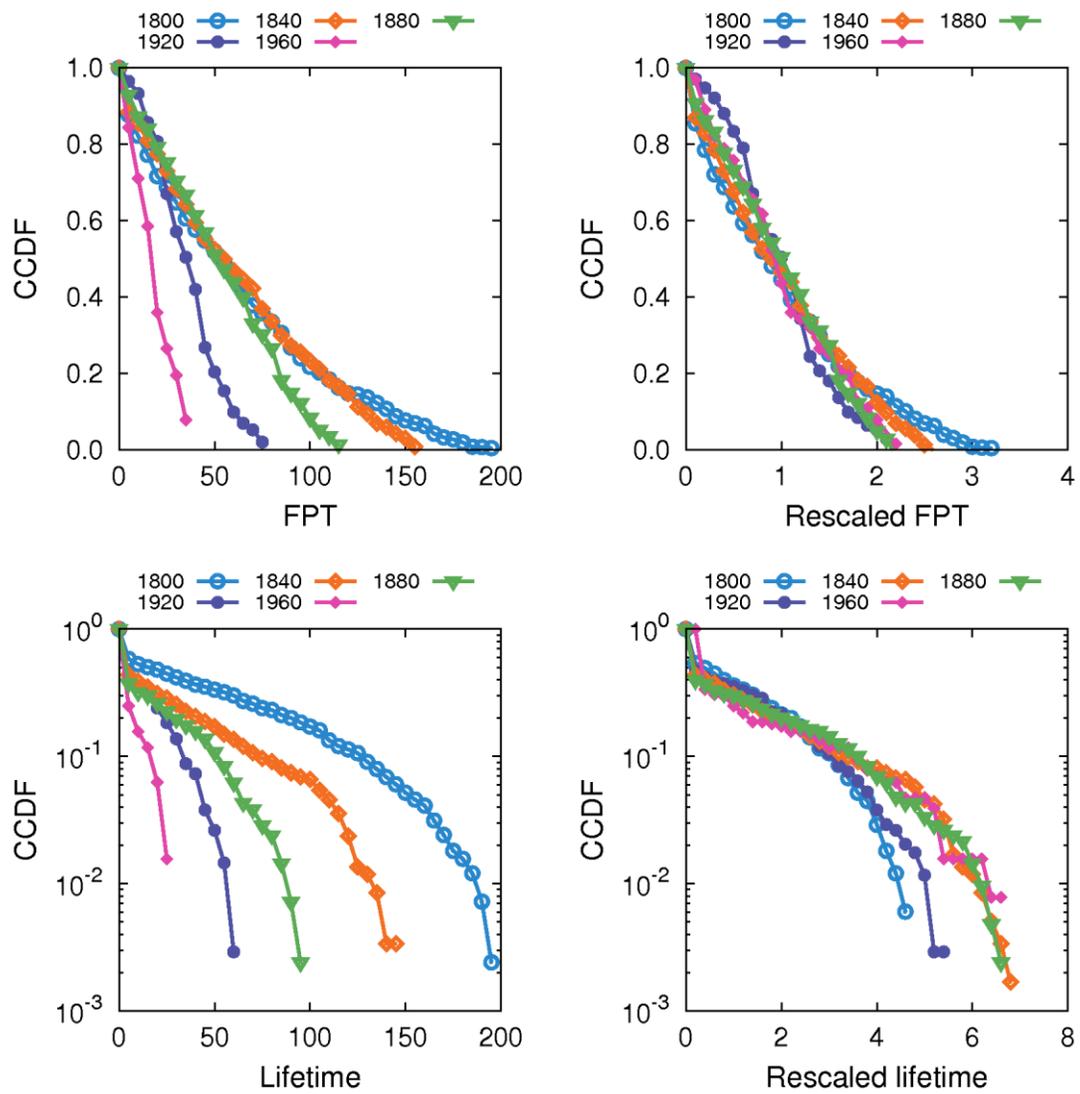

**Figure S7**. Complementary cumulative distribution functions (CCDFs) of FPT, lifetime, and their rescaled values for each set of all type-I 1-grams from the same year of birth.

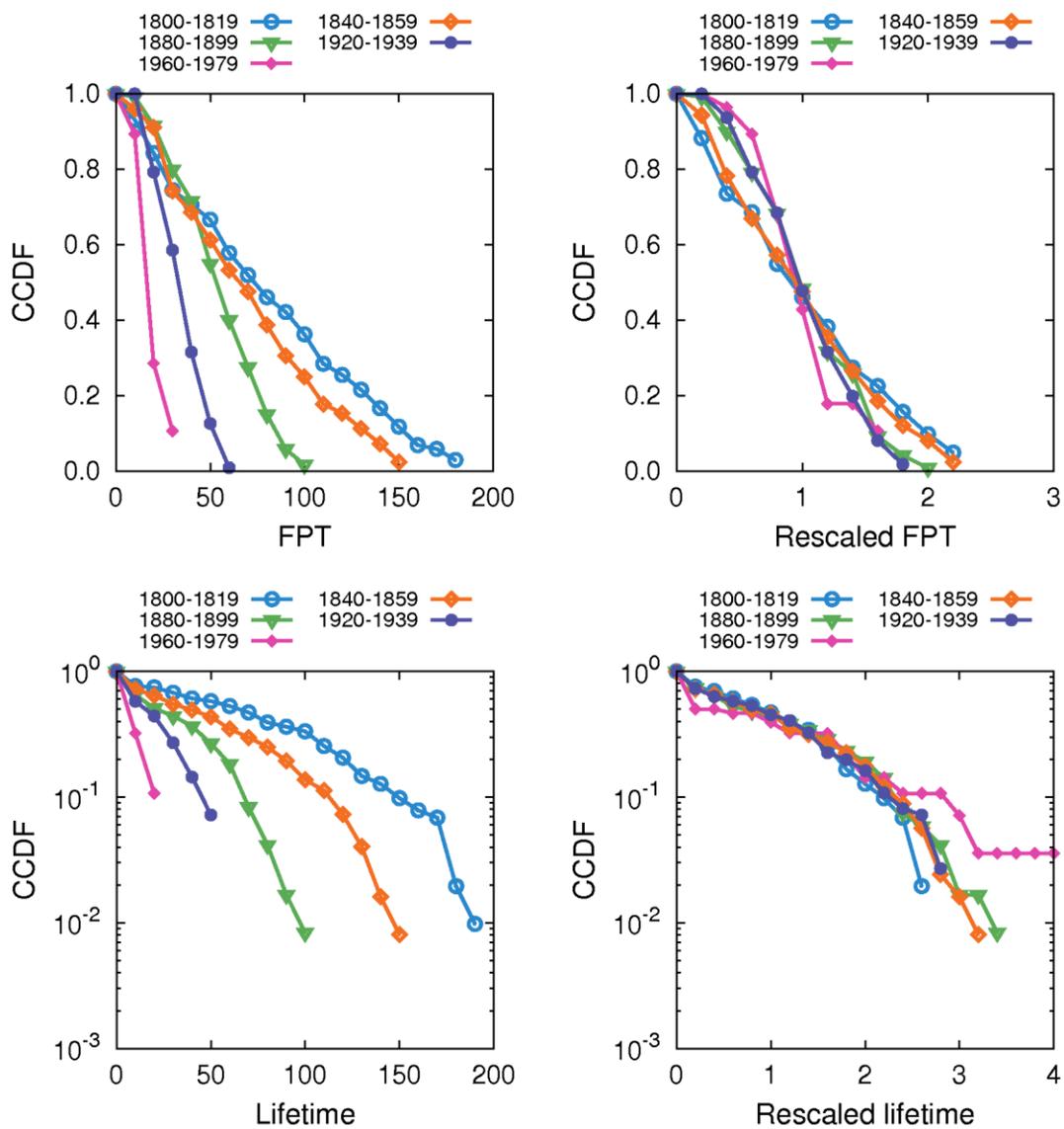

**Figure S8**. Complementary cumulative distribution functions (CCDFs) of FPT, lifetime, and their rescaled values for each set of type-I scientific words from the same year of birth.

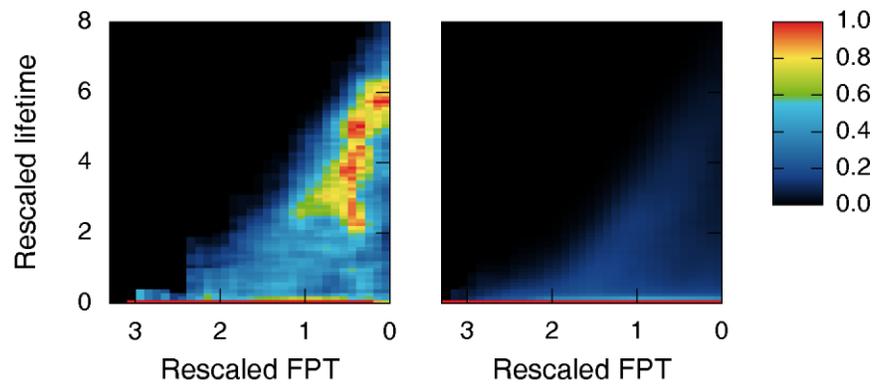

**Figure S9.** Density plot between FPT and lifetime of type-I scientific words (left) and all type-I 1-grams (right). Coloured according to adjusted density, following the scale bar on the rightmost side. $b_x = 0.4$, $b_y = 0.8$, $k_x = 4$, and $k_y = 4$ (see Supplementary Methods).

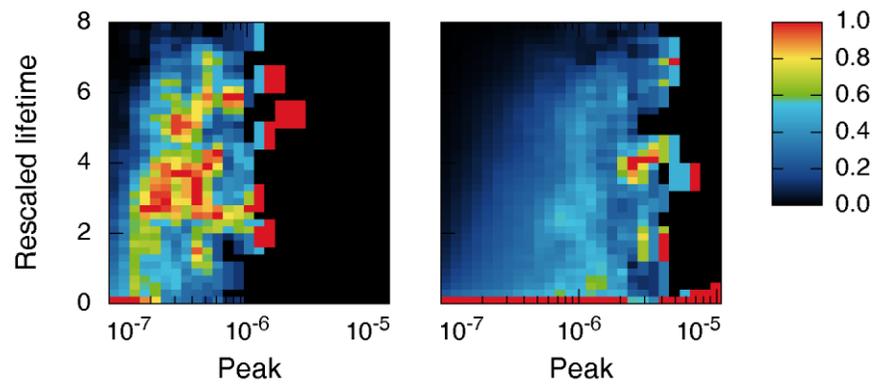

**Figure S10.** Density plot between peak and lifetime of type-I scientific words (left) and all type-I 1-grams (right). Coloured according to adjusted density, following the scale bar on the rightmost side. $b_x = 10^{-7}$, $b_y = 0.4$, $k_x = 4$, $k_y = 4$, and $i_x = 2$ (see Supplementary Methods).

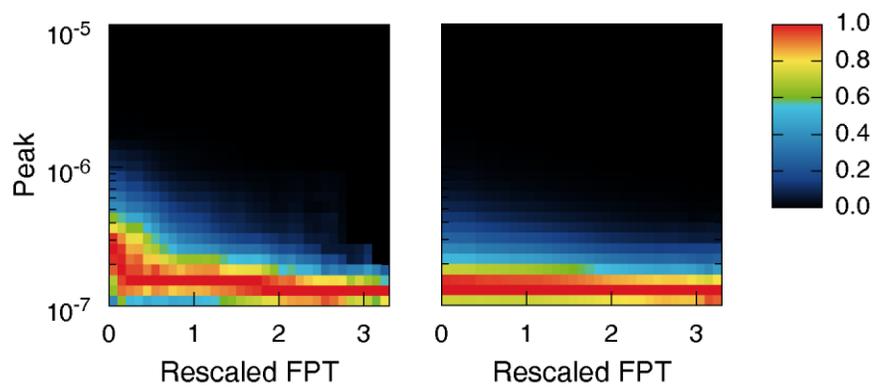

**Figure S11.** Density plot between FPT and peak of type-I scientific words (left) and all type-I 1-grams (right). Coloured according to adjusted density, following the scale bar on the rightmost side. $b_x = 0.8$, $b_y = 10^{-7}$, $k_x = 4$, $k_y = 4$, and $i_y = 2$ (see Supplementary Methods).

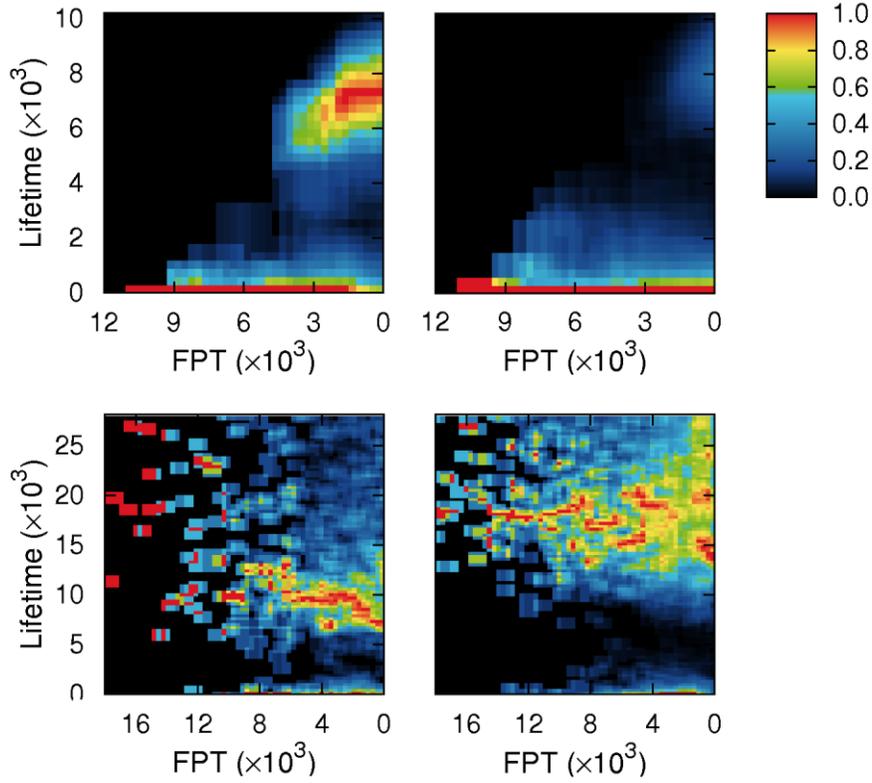

**Figure S12.** Density plot between FPT and lifetime from the simulation with the power-law fitness distribution ($\gamma = 2.0$, $\beta = 1/4$, $N = 4{,}096$, $L = 10$, $p_{\text{ER}} = 0.1024$, $\rho = 0.2$, $\alpha = 0.0001$, $f_c = 0.00025$). Coloured according to adjusted density, following the scale bar on the rightmost side ($b_x = 2{,}400$, $b_y = 2{,}400$, $k_x = 12$, $k_y = 12$ for the top panel and $b_x = 1{,}600$, $b_y = 1{,}600$, $k_x = 8$, $k_y = 8$ for the bottom panel; see Supplementary Methods). The left panels are for scientific items and the right panels are for the rest. Each top panel is a magnification of the region for FPT + lifetime < 12,000 in the bottom panel, representing the type-I case.

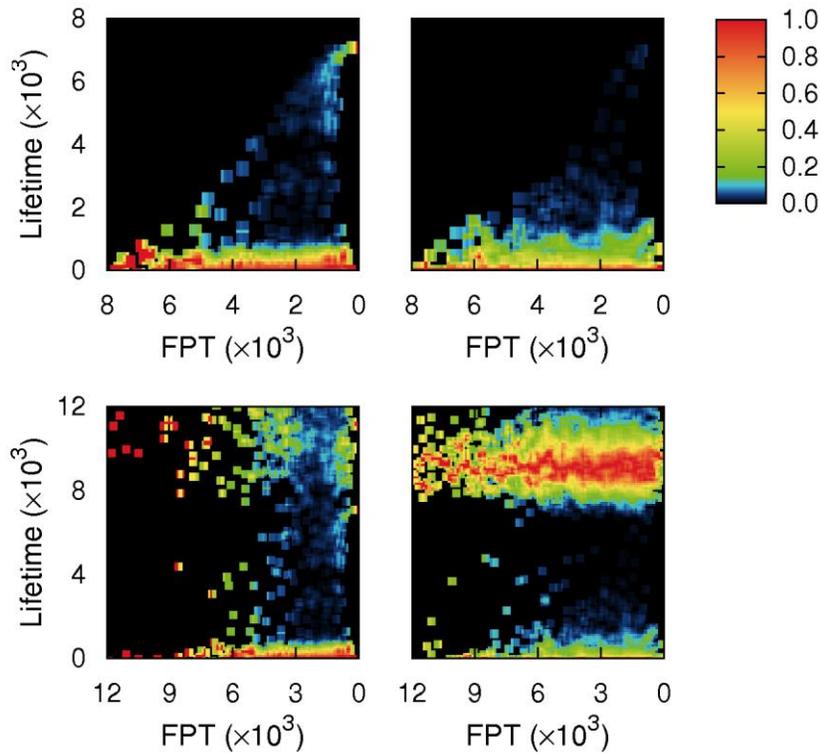

**Figure S13.** Density plot between FPT and lifetime from the simulation with the Gaussian fitness distribution ($\sigma = 0.1$, $N = 4{,}096$, $L = 10$, $p_{\mathrm{ER}} = 0.0064$, $\rho = 0.15$, $\alpha = 0.0001$, $f_c = 0.00025$). Coloured according to adjusted density, following the scale bar on the rightmost side ($b_x = 400$, $b_y = 400$, $k_x = 4$, $k_y = 4$; see Supplementary Methods). The left panels are for scientific items and the right panels are for the rest. Each top panel is a magnification of the region for FPT + lifetime < 8,400 in the bottom panel, representing the type-I case.

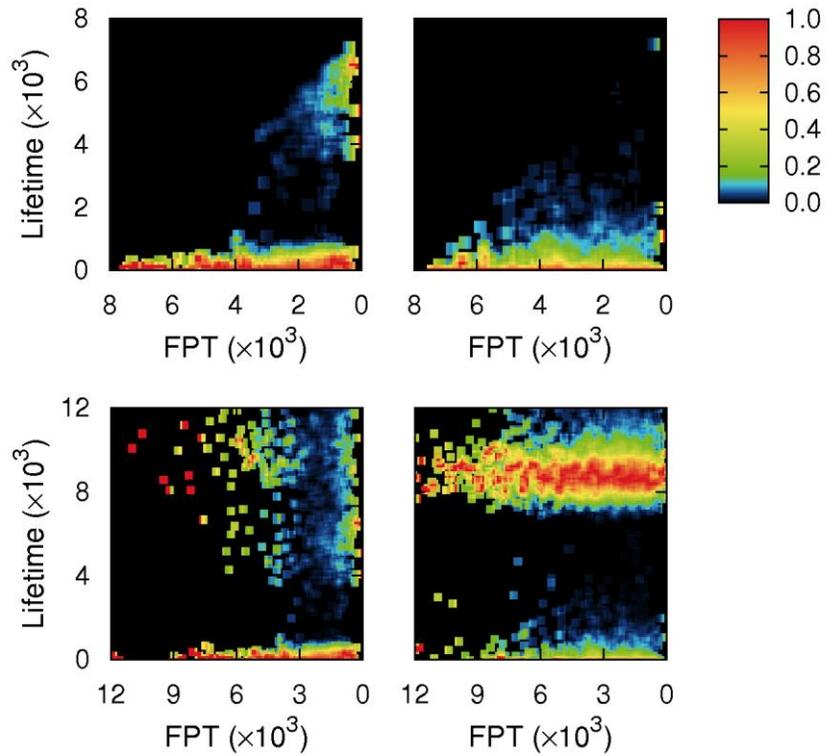

**Figure S14.** Density plot between FPT and lifetime from the simulation with the Dirac delta distribution of fitness ($N = 4{,}096$, $L = 10$, $p_{ER} = 0.0064$, $\rho = 0.15$, $\alpha = 0.0001$, $f_c = 0.00025$). Coloured according to adjusted density, following the scale bar on the rightmost side ($b_x = 400$, $b_y = 400$, $k_x = 4$, $k_y = 4$; see Supplementary Methods). The left panels are for scientific items and the right panels are for the rest. Each top panel is a magnification of the region for FPT + lifetime < 8,000 in the bottom panel, representing the type-I case.

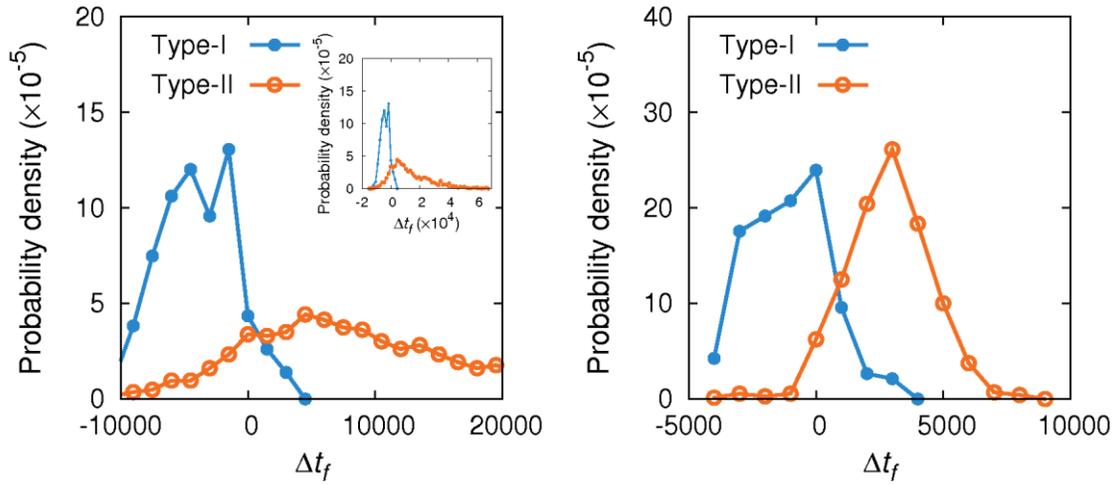

**Figure S15.** Probability distributions of $\Delta t_f = t_f' - t_f$ for type-I and type-II scientific items, where $t_f'$ ($t_f$) of each item is the last time that the frequency of the item outside (inside) the scientific community fell below $f_c$. The left panel is for the power-law fitness distribution (inset for the full range of $\Delta t_f$; $\gamma = 2.0$, $\beta = 1/4$, $N = 4{,}096$, $L = 10$, $p_{ER} = 0.1024$, $\rho = 0.2$, $\alpha = 0.0001$, $f_c = 0.00025$) and the right panel is for the Gaussian fitness distribution ($\sigma = 0.1$, $N = 4{,}096$, $L = 10$, $p_{ER} = 0.0064$, $\rho = 0.15$, $\alpha = 0.0001$, $f_c = 0.00025$).

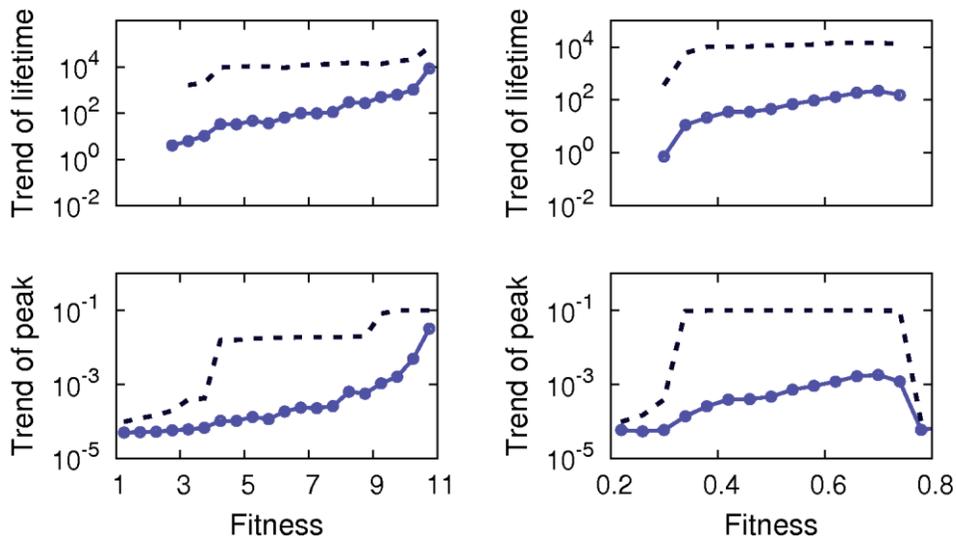

**Figure S16.** Dependency of lifetime and peak of scientific items (for all three types) on fitness. The left panels are for the power-law fitness distribution ($\gamma = 2.0$, $\beta = 1/4$, $N = 4{,}096$, $L = 10$, $p_{ER} = 0.1024$, $\rho = 0.2$, $\alpha = 0.0001$, $f_c = 0.00025$) and the right panels are for the Gaussian fitness distribution ($\sigma = 0.1$, $N = 4{,}096$, $L = 10$, $p_{ER} = 0.0064$, $\rho = 0.15$, $\alpha = 0.0001$, $f_c = 0.00025$). Lifetime of type-III was treated as zero for this analysis. For the lifetime and peak at each level of fitness, filled circles represent the averages and dashed lines represent the upper 0.1% of lifetime and peak.

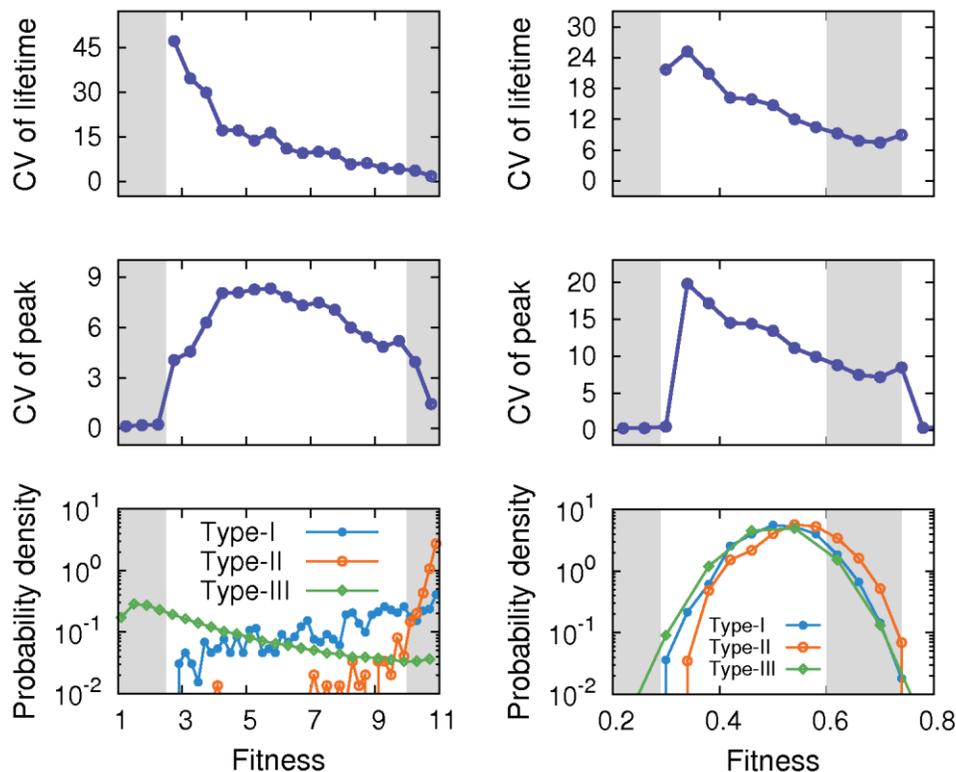

**Figure S17.** Variability of lifetime and peak of scientific items (for all three types) across fitness. The left panels are for the power-law fitness distribution ($\gamma = 2.0$, $\beta = 1/4$, $N = 4{,}096$, $L = 10$, $p_{ER} = 0.1024$, $\rho = 0.2$, $\alpha = 0.0001$, $f_c = 0.00025$) and the right panels are for the Gaussian fitness distribution ($\sigma = 0.1$, $N = 4{,}096$, $L = 10$, $p_{ER} = 0.0064$, $\rho = 0.15$, $\alpha = 0.0001$, $f_c = 0.00025$). Lifetime of type-III was treated as zero for this analysis. The upper two panels show the coefficients of variation (CVs) of lifetime and peak for each level of fitness, and the bottom panel shows the probability distributions of fitness for different types. The shaded areas indicate the ranges of fitness toward which the distributions of type-II and type-III are biased (type-II for high fitness and type-III for low fitness). In the left shaded areas, CVs of lifetime are ill-defined because all items in the areas belong to type-III with zero lifetime; the variability of lifetime in these areas can be viewed as effectively zero.

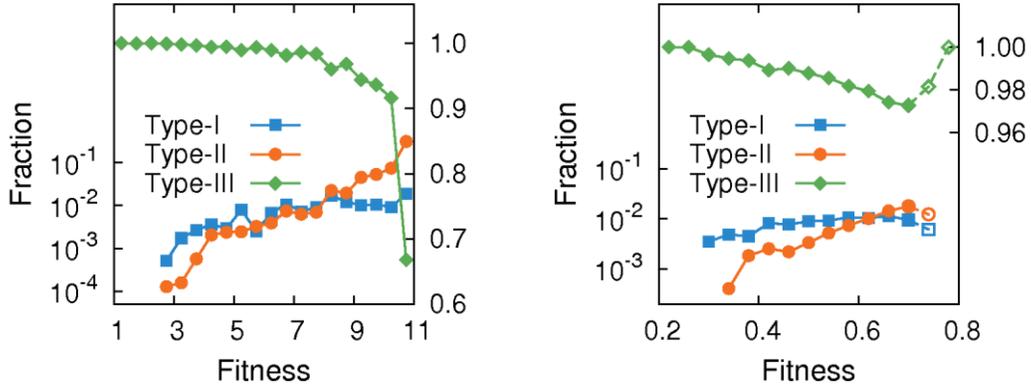

**Figure S18.** Ratios of different types to the total at each level of fitness for scientific items. The left panel is for the power-law fitness distribution ($\gamma = 2.0$, $\beta = 1/4$, $N = 4{,}096$, $L = 10$, $p_{ER} = 0.1024$, $\rho = 0.2$, $\alpha = 0.0001$, $f_c = 0.00025$) and the right panel is for the Gaussian fitness distribution ($\sigma = 0.1$, $N = 4{,}096$, $L = 10$, $p_{ER} = 0.0064$, $\rho = 0.15$, $\alpha = 0.0001$, $f_c = 0.00025$). For types-I and -II, plotted in logarithmic scale (left axis). For type-III, plotted in linear scale (right axis). The fractions of type-I and type-II tend to increase as fitness increases, but the slope is steeper for type-II. The fraction of type-III is very high in all ranges of fitness, but slightly increases as fitness decreases (in the right panel, open circles with dashed lines do not have a statistically-meaningful number of items).

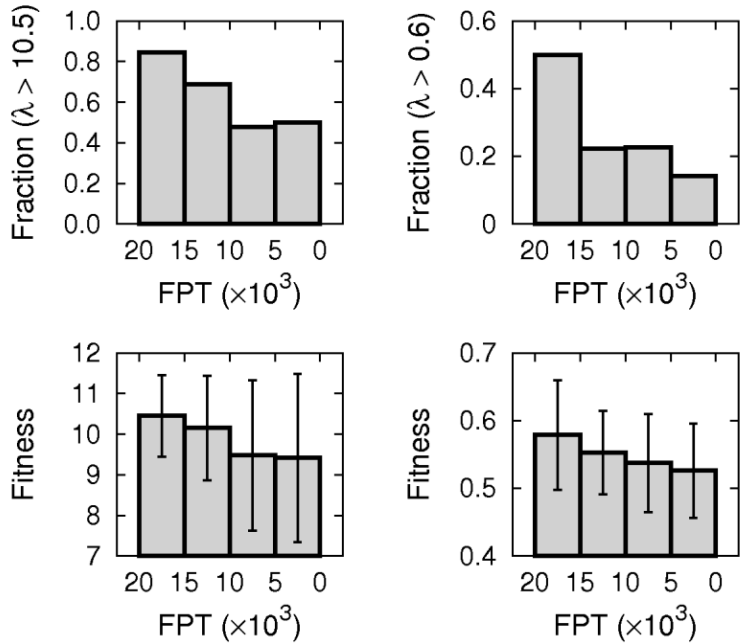

**Figure S19.** Relation between FPT and fitness of types-I and -II scientific items. We checked (left) the case for the power-law distribution of fitness ($\gamma = 2.0$, $\beta = 1/4$, $N = 4{,}096$, $L = 10$, $p_{ER} = 0.1024$, $\rho = 0.2$, $\alpha = 0.0001$, $f_c = 0.00025$) and (right) that for the Gaussian distribution of fitness ($\sigma = 0.1$, $N = 4{,}096$, $L = 10$, $p_{ER} = 0.0064$, $\rho = 0.15$, $\alpha = 0.0001$, $f_c = 0.00025$). The top panels show the fractions of fitness $> 10.5$ (left) and $> 0.6$ (right) for each range of FPT. The bottom panels show the averages and the standard deviations of fitness for each range of FPT.

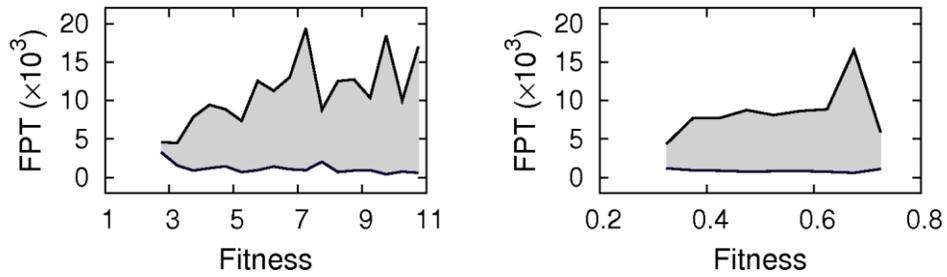

**Figure S20.** Extent of FPT for types-I and -II scientific items. We checked (left) the case for the power-law distribution of fitness ($\gamma = 2.0$, $\beta = 1/4$, $N = 4{,}096$, $L = 10$, $p_{\text{ER}} = 0.1024$, $\rho = 0.2$, $\alpha = 0.0001$, $f_c = 0.00025$) and (right) that for the Gaussian distribution of fitness ($\sigma = 0.1$, $N = 4{,}096$, $L = 10$, $p_{\text{ER}} = 0.0064$, $\rho = 0.15$, $\alpha = 0.0001$, $f_c = 0.00025$). Shaded area is between the upper and lower 1%s in FPT for each fitness.

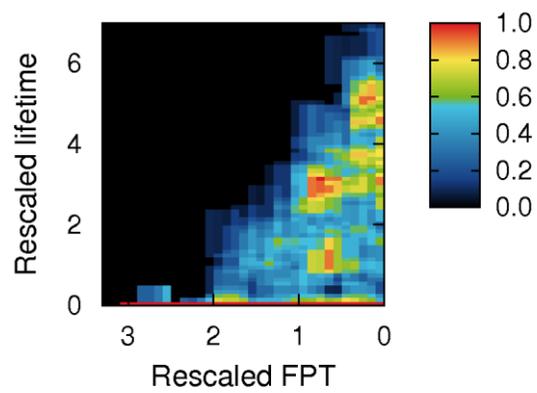

**Figure S21.** Density plot between FPT and lifetime of type-I food and nutritional words without those analysed for scientific evolution. $b_x = 0.6$, $b_y = 1.2$, $k_x = 6$, and $k_y = 12$ (see Supplementary Methods).

# Supplementary Tables

**Table S1.** List of sources where we collected scientific and technological words.

| Source | Period | Source | Period |
|---|---|---|---|
| *AccessScience* | 2000 – 2012 | *United States Patent and Trademark Office* | 1920 – 1979 |
| *Philosophical Transactions of the Royal Society* | 1665 – 1887 | *Philosophical Transactions A, B* | 1887 – 2012 |
| *Science* | 1880 – 2012 | *Nature* | 1869 – 2012 |
| *Proceedings of the National Academy of Sciences of the United States of America* | 1914 – 2012 | *Physical Review* | 1893 – 1969 |
| *Physical Review A, B, C, D* | 1970 – 2009 | *Physical Review E* | 1993 – 2009 |
| *Physical Review Letters* | 1958 – 2009 | *Review of Modern Physics* | 1929 – 2009 |
| *Physical Review Special Topics – Physics Education Research* | 2005 – 2009 | *Physical Review Special Topics – Accelerators and Beams* | 1998 – 2009 |

**Table S2.** List of scientific words predicted to be type-II, which first passed the frequency of $2.0\times10^{-6}$ in the years between 2000 and 2008 (the exact years are recorded in the second column). From Figure 1b, the chance of being type-II is estimated to be 97.1%. We assigned a category and a context of use to each word according to its common usage.

| Stem | First crossing year | Corresponding word | Category | Context of use | Description |
|---|---|---|---|---|---|
| p53 | 2006 | p53 | Bio | Cancer | Tumour suppressor. |
| cortisol | 2001 | cortisol | Med | Drug | Steroid hormone. It is used as an immunosuppressive drug. |
| nanoparticl | 2004 | nanoparticle | Nano | | Small particle in a nanoscale. |
| nanotechnolog | 2006 | nanotechnology | Nano | | Technology to manipulate atomic- or molecular-level objects. |
| nanotub | 2008 | nanotube | Nano | | Nanometer scale tube-like object, e.g., carbon nanotube. |
| tsunami | 2006 | tsunami | Geo | | A series of water waves in large scale. |
| dataset | 2001 | dataset | Etc | IT | Collection of data. |

**Table S3.** Scientific words predicted to be type-II. For the frequency of $1.5 \times 10^{-6}$ first passed in the years between 2000 and 2008, they are listed in a similar way to Table S2.

| Stem | First crossing year | Corresponding word | Category | Context of use | Description |
|---|---|---|---|---|---|
| isoform | 2006 | isoform | Bio | Disease | Different forms of the same protein. It is related to Mad Cow Disease. |
| bioethic | 2007 | bioethics | Bio | | Ethics related to biological and medical issues. |
| biofilm | 2000 | biofilm | Bio | | Group of microorganisms sticking to each other on a surface. |
| biomark | 2006 | biomarker | Med | Disease | Measurable characteristics that sense a sign of disease. |
| thrombo-cytopenia | 2002 | thrombo-cytopenia | Med | Disease | Decrease of platelets in blood. |
| osteoarthr | 2007 | osteoarthritis | Med | Aging, Disease | Degenerative joint disease. |
| nanotechnolog | 2006 | nanotechnology | Nano | | Technology to manipulate atomic- or molecular-level objects. |
| nanoparticl | 2004 | nanoparticle | Nano | | Small particle in a nanoscale. |
| nanotub | 2004 | nanotube | Nano | | Nanometer scale tube-like object, e.g., carbon nanotube. |
| tsunami | 2005 | tsunami | Geo | | A series of water waves in large scale. |
| holocen | 2007 | Holocene | Geo | | Geological epoch that began at the end of the Pleistocene and continues to the present. |

**Table S4.** Scientific words predicted to be type-II. For the frequency of $1.0\times10^{-6}$ first passed in the years between 2000 and 2008, they are listed in a similar way to Table S2.

| Stem | First crossing year | Corresponding word | Category | Context of use | Description |
|---|---|---|---|---|---|
| vegf | 2006 | VEGF | Bio | Cancer | Growth factor that stimulates blood vessel formation. Cancer related. |
| transferrin | 2002 | transferrin | Bio | Disease | Blood plasma glycoprotein that controls the level of free iron. |
| biofuel | 2008 | biofuel | Bio | Energy, Environment | Fuel produced from living organisms. |
| reuptak | 2006 | reuptake | Bio | Neurological diseases and disorders | Reabsorption of a neurotransmitter by a pre-synaptic neuron. |
| cd8 | 2007 | CD8 | Bio | | Co-receptor for the T cell receptor, mainly expressed in cytotoxic T cells. |
| cyanobacteria | 2000 | cyanobacteria | Bio | | Bacteria capable of photosynthesis. |
| nephropathi | 2004 | nephropathy | Med | Aging, Disease | Kidney disease. |
| biomark | 2006 | biomarker | Med | Disease | Measurable characteristics that sense a sign of disease. |
| neurologist | 2007 | neurologist | Med | Neurological diseases and disorders | Physician who specializes in neurology. |
| biomateri | 2006 | biomaterial | Med | | Any object that interacts with biological systems. |
| biosensor | 2007 | biosensor | Med | | Device for the detection of biological components. |
| detoxif | 2001 | detoxification | Med | | Removal of toxic substances from a living body. |
| microsystem | 2001 | microsystem | Nano | | Miniaturized device for non-electronic function. |
| nanoparticl | 2003 | nanoparticle | Nano | | Small particle in a nanoscale. |

| Stem | First crossing year | Corresponding word | Category | Context of use | Description |
|---|---|---|---|---|---|
| nanotub | 2003 | nanotube | Nano | | Nanometer scale tube-like object, e.g., carbon nanotube. |
| nanotechnolog | 2005 | nanotechnology | Nano | | Technology to manipulate atomic- or molecular-level objects. |
| nanostructur | 2006 | nanostructure | Nano | | Object of molecular or microscopic structure. |

**Table S5.** Scientific words that first passed the frequency of $5.0 \times 10^{-7}$ in the years between 2000 and 2008. They are listed in a similar way to Table S2. Although the list includes the words with a rather low chance of being type-II (78.5%) compared with Tables S2–S4 ($\geq$ 90.4%), we provide it for supplementary purposes.

| Stem | First crossing year | Corresponding word | Category | Context of use | Description |
|---|---|---|---|---|---|
| h2o2 | 2006 | $H_2O_2$ | Bio | Aging | Simplest peroxide. Highly reactive oxygen species. |
| polyphenol | 2006 | polyphenol | Bio | Aging | Molecule with large multiples of phenol structural units. Antioxidant effect. |
| brca1 | 2007 | BRCA1 | Bio | Cancer | Breast cancer type 1 susceptibility protein. |
| isoflavon | 2002 | isoflavone | Bio | Cancer | A class of organic compounds, often naturally occurring, related to the isoflavonoids. |
| microenviron | 2007 | micro-environment | Bio | Cancer | Small-scale environment around cells. |
| vegf | 2001 | VEGF | Bio | Cancer | Growth factor that stimulates blood vessel formation. Cancer related. |
| proteasom | 2004 | proteasome | Bio | Cancer, Neurological diseases and disorders | Protein complex that degrades unneeded or damaged proteins. Related to cervical cancer and cystic fibrosis. |
| cardiomyocyt | 2007 | cardiomyocyte | Bio | Disease | Heart muscle. Related to cardiomyopathy. |
| metallo-proteinas | 2005 | metallo-proteinase | Bio | Disease | Protease enzyme with catalytic activity involving a metal. Used to treat periodontal disease. |
| ubiquitin | 2002 | ubiquitin | Bio | Neurological diseases and disorders | Small regulatory protein leading to protein degradation. Related to Alzheimer's and Parkinson's diseases. |
| cannabinoid | 2002 | cannabinoid | Bio | Neurological diseases and disorders | Chemical compound acting on cannabinoid receptors on cells that repress neurotransmitter release in the brain. |

| Stem | First crossing year | Corresponding word | Category | Context of use | Description |
|---|---|---|---|---|---|
| microglia | 2006 | microglia | Bio | Neurological diseases and disorders | Macrophages in the brain. Related to Alzheimer's and Parkinson's diseases. |
| neurogenesi | 2006 | neurogenesis | Bio | Neurological diseases and disorders | Process to generate neurons. |
| hydrogel | 2000 | hydrogel | Bio | Tissue engineering | Hydrophilic polymer network. |
| biofuel | 2007 | biofuel | Bio | Energy, Environment | Fuel produced from living organisms. |
| archaea | 2007 | archaea | Bio | | One of the three domains of life. |
| aspartam | 2001 | aspartame | Bio | | Non-saccharide, sugar substitute. |
| biomolecul | 2006 | biomolecule | Bio | | Any molecule produced by a living organism. |
| fluorophor | 2006 | fluorophore | Bio | | Fluorescent chemical compound. |
| immunohisto-chemistri | 2006 | Immunohisto-chemistry | Bio | | Process of detecting antigens in cells. Used for diagnosis. |
| ligas | 2006 | ligase | Bio | | Enzyme for the joining of two large molecules. |
| luciferas | 2006 | luciferase | Bio | | Enzyme for bioluminescence. |
| paclitaxel | 2005 | paclitaxel | Med | Cancer | Mitotic inhibitor used in cancer chemotherapy. |
| hemo-chromatosi | 2000 | hemo-chromatosis | Med | Disease | Iron overload in the body. |
| metformin | 2007 | metformin | Med | Drug | Antidiabetic drug. |

| Stem | First crossing year | Corresponding word | Category | Context of use | Description |
|---|---|---|---|---|---|
| prodrug | 2007 | prodrug | Med | Drug | Drug initially administered in an inactive form, later converted to its active form in the body. |
| meth-amphetamin | 2004 | meth-amphetamine | Med | Drug, Neurological diseases and disorders | Narcotic to treat attention deficit hyperactivity disorder (ADHD). |
| valproat | 2007 | valproate | Med | Drug, Neurological diseases and disorders | Drug to treat epilepsy and bipolar disorder. |
| phytolith | 2006 | phytolith | Geo | | Silica from plants. Used for archaeological and paleoenvironmental research. |
| plasmon | 2006 | plasmon | Physics | | Quantum of plasma oscillation. |
| supersymmetri | 2000 | supersymmetry | Physics | | Symmetry that relates two basic classes of particles, bosons and fermions. |
| dendrim | 2006 | dendrimer | Nano | | Repetitively branched molecules. Potential use for drug and gene delivery. |
| mesopor | 2004 | mesopore | Nano | | Material with a nanoscale pores. |
| nanocryst | 2003 | nanocrystal | Nano | | Crystal structure in a nanoscale. |
| nanomateri | 2007 | nanomaterial | Nano | | Material in a nanoscale. |
| nanoparticl | 2001 | nanoparticle | Nano | | Small particle in a nanoscale. |
| nanostructur | 2002 | nanostructure | Nano | | Object of molecular or microscopic structure. |
| nanotechnolog | 2003 | nanotechnology | Nano | | Technology to manipulate atomic- or molecular-level objects. |

| Stem | First crossing year | Corresponding word | Category | Context of use | Description |
|---|---|---|---|---|---|
| nanotub | 2002 | nanotube | Nano | | Nanometer scale tube-like object, e.g., carbon nanotube. |
| nanowir | 2004 | nanowire | Nano | | Nanometer scale wire-like object. |
| tio2 | 2005 | $TiO_2$ | Nano | | Compound used to make inorganic nanotubes. |
| lightwav | 2002 | LightWave | IT | | 3D graphics tool to make movies and computer games. |
| *spywar* | 2005 | spyware | IT | | Software to steal information from computers. |
| radiofrequ | 2000 | radiofrequency | Etc. | IT | Frequency of about 3kHz to 300GHz. Noticed for recent technologies such as radio-frequency identification (RFID). |
| tribolog | 2001 | tribology | Etc. | Nano, Bio, Environmental | Study of interacting surfaces in relative motion. |

**Table S6.** *Z*-scores and *P*-values for the fraction of type-II for scientific words having each level of frequency passed between 1800 and 1919.

| Frequency level | Z-score | P-value |
| --- | --- | --- |
| $2.0\times10^{-6}$ | 5.93 | $3.04\times10^{-9}$ |
| $1.5\times10^{-6}$ | 8.02 | $1.07\times10^{-15}$ |
| $1.0\times10^{-6}$ | 9.25 | $2.28\times10^{-20}$ |
| $5.0\times10^{-7}$ | 10.25 | $1.23\times10^{-24}$ |
| $1.0\times10^{-7}$ | 9.15 | $5.85\times10^{-20}$ |
| $1.0\times10^{-8}$ | 4.59 | $4.41\times10^{-6}$ |
| $1.0\times10^{-9}$ | 0.46 | 0.64 |

**Table S7.** Scientific words of late bloomer candidates (type-II with rescaled FPT ≥ 2.0). Excluded are those subject to dating or OCR errors and to non-scientific use.

| Stem | Corresponding word | FPT (years) | Year of birth | Rescaled FPT | Description |
|---|---|---|---|---|---|
| eudicot | eudicots | 205 | 1802 | 2.96 | Monophyletic clade of flowering plant. |
| cuprat | cuprate | 188 | 1804 | 2.77 | Material containing copper anions. |
| retinoid | retinoid | 173 | 1809 | 2.73 | Vitamin A. |
| toxicologist | toxicologist | 166 | 1809 | 2.62 | Professional who specializes in the poisoning of living organisms. |
| megafauna | megafauna | 82 | 1926 | 2.61 | Large or giant animal in terrestrial zoology. |
| microgel | microgel | 75 | 1932 | 2.61 | Cross-linked three-dimensional polymer networks swollen in a solvent. |
| phosphopeptid | phosphopeptide | 54 | 1950 | 2.60 | Phosphorylated peptide. |
| niobat | niobate | 143 | 1846 | 2.56 | Salt containing an anionic grouping of niobium and oxygen. |
| micro-architectur | micro-architecture | 53 | 1950 | 2.55 | Way to implement an instruction set architecture (ISA) on a computer processor. |
| speleothem | speleothem | 52 | 1953 | 2.54 | Mineral deposit formed in a cave. |
| endosymbiont | endosymbiont | 68 | 1939 | 2.53 | Organism living within the body or cells of another organism. |
| nematolog | nematology | 91 | 1914 | 2.53 | Study of nematodes. |
| superoxid | superoxide | 170 | 1804 | 2.5 | Compound including the superoxide anion. |

| Stem | Corresponding word | FPT (years) | Year of birth | Rescaled FPT | Description |
| --- | --- | --- | --- | --- | --- |
| metalloproteas | metalloprotease | 48 | 1959 | 2.48 | Protease enzyme involving a metal in its catalytic activity. |
| trans-glutaminas | trans-glutaminase | 48 | 1958 | 2.48 | Enzyme that links an amine group and glutamine. |
| steatosi | steatosis | 165 | 1835 | 2.47 | Abnormal retention of lipids within a cell. |
| autophagi | autophagy | 165 | 1825 | 2.46 | Intracellular degradation of unnecessary or dysfunctional cellular components. |
| agaros | agarose | 149 | 1819 | 2.45 | Polysaccharide material extracted from seaweed. |
| allelopathi | allelopathy | 68 | 1931 | 2.45 | Biochemical interactions between organisms to affect their growth, survival, and reproduction. |
| vasculopathi | vasculopathy | 53 | 1949 | 2.44 | Disorder of blood vessels. |
| gapdh | GAPDH | 47 | 1959 | 2.43 | Enzyme involved in glycolysis. |
| spallat | spallation | 147 | 1810 | 2.43 | Ejection of fragments from a material by impact or stress. |
| lipodystrophi | lipodystrophy | 94 | 1910 | 2.42 | Abnormal or degenerative condition of adipose tissues. |
| cryptolog | cryptology | 163 | 1805 | 2.41 | Technique for secure communication in the presence of third parties. |
| glucokinas | glucokinase | 57 | 1948 | 2.39 | Enzyme to phosphorylate glucose. |
| discoideum | *Dictyostelium discoideum* | 163 | 1812 | 2.35 | Soil-living amoeba. |
| xenotransplant | xenotransplant | 61 | 1937 | 2.35 | Animal tissue or organ transplant in a human patient. |

| Stem | Corresponding word | FPT (years) | Year of birth | Rescaled FPT | Description |
|---|---|---|---|---|---|
| isoflavon | isoflavone | 70 | 1928 | 2.34 | A class of organic compounds, often naturally occurring, related to the isoflavonoids. |
| azoospermia | azoospermia | 128 | 1870 | 2.33 | Medical condition of a man without any measurable level of sperms. |
| selenoprotein | selenoprotein | 105 | 1902 | 2.32 | Protein including a selenocysteine. |
| thermotherapi | thermotherapy | 135 | 1873 | 2.32 | Application of heat to the body for health and medical purpose. |
| primatologist | primatologist | 74 | 1927 | 2.31 | Professional who studies primates. |
| biofuel | biofuel | 58 | 1940 | 2.28 | Fuel produced from living organisms. |
| ecosystem | ecosystem | 136 | 1817 | 2.28 | Community of living organisms in the environment. |
| bioenergi | bioenergy | 68 | 1928 | 2.27 | Renewable energy from biological sources. |
| dyslipidemia | dyslipidemia | 55 | 1946 | 2.27 | Abnormal amount of lipids in the blood. |
| microsensor | microsensor | 54 | 1948 | 2.26 | Sensing device of small size. |
| pervapor | pervaporation | 74 | 1918 | 2.26 | Separation of liquid mixtures through a non-porous or porous membrane. |
| audiometri | audiometry | 141 | 1809 | 2.23 | Branch of audiology for measurements of hearing acuity. |
| polyomavirus | *Polyomavirus* | 43 | 1959 | 2.22 | Oncogenic virus. |
| recombinas | recombinase | 43 | 1959 | 2.22 | Enzyme for genetic recombination. |

| Stem | Corresponding word | FPT (years) | Year of birth | Rescaled FPT | Description |
|---|---|---|---|---|---|
| transesterif | transe-sterification | 59 | 1938 | 2.21 | Exchange of organic groups in an ester and an alcohol. |
| asteracea | Asteraceae | 147 | 1835 | 2.2 | Aster, daisy or sunflower family. |
| desulfovibrio | *Desulfovibrio* | 57 | 1937 | 2.20 | Sulfate-reducing bacteria. |
| microcav | microcave | 52 | 1944 | 2.2 | Structure to confine light to small volumes by resonant recirculation. |
| midgut | midgut | 131 | 1810 | 2.17 | Part of the embryo to develop into the intestines. |
| paleo-anthropolog | paleo-anthropology | 91 | 1907 | 2.17 | Study of ancient humans as found in fossils. |
| theropod | theropod | 103 | 1886 | 2.14 | Suborder of dinosaurs. |
| homoplasi | homoplasy | 126 | 1871 | 2.12 | Independent evolution of similar features in different taxa. |
| aerogel | aerogel | 69 | 1918 | 2.11 | Gel of which liquid component has been replaced by gas. |
| retinol | retinol | 128 | 1842 | 2.11 | Vitamin A. |
| polyhedra | polyhedra | 145 | 1802 | 2.09 | Three dimensional geometrical object with flat faces and straight edges. |
| bloodstream | bloodstream | 124 | 1817 | 2.08 | Blood flow through the circulatory system. |
| microcentrifug | microcentrifuge | 70 | 1922 | 2.08 | Equipment that spins liquid samples at high speed. |
| phytas | phytase | 90 | 1909 | 2.05 | Phosphatase enzyme that hydrolyses phytic acid. |

| Stem | Corresponding word | FPT (years) | Year of birth | Rescaled FPT | Description |
|---|---|---|---|---|---|
| capsaicin | capsaicin | 106 | 1877 | 2.02 | Component of chili peppers. |
| angiopathi | angiopathy | 124 | 1852 | 2.00 | Disease of the blood vessels. |

# Supplementary Methods

## 1. Dataset

### 1.1. *Google Books Ngram Corpus*

We obtained the annual data of *n*-gram counts contained in the English section of the *Google Books Ngram Corpus* Version 2 which spans 8,116,746 books published over the last five centuries [1]. A 1-gram is a string of characters uninterrupted by a space, e.g., a word or number. An *n*-gram is a sequence of 1-grams, e.g., *n*=3 for a phrase with three words. We here focus on 1-grams for simplicity of analysis. A sample of the data is given below (from the data file "googlebooks-eng-all-1gram-20120701-w.csv"):

| work | 2000 | 7285922 | 100673 |
| work_VERB | 2000 | 1848009 | 93981 |
| work_NOUN | 2000 | 5377995 | 99002 |

The first row shows that in 2000, the word "work" occurred 7,285,922 times in 100,673 different books. The second and third rows show that in the same year, the word "work" occurred 1,848,009 times as a verb, and 5,377,995 times as a noun. Relative frequency of a 1-gram is defined as the number of occurrences of the 1-gram in a given year divided by the total number of 1-grams in that year.

### 1.2. Preprocessing

In order to treat different inflectional forms (e.g., singular and plural) of the same word stem as equivalent in their essential meaning, we integrated such forms systematically by *Porter Stemming Algorithm* [2] when computing the 1-gram frequency. We also limited our analysis to data in the years between 1800 and 2008 because the amount of data before 1800 is not sufficient to obtain statistically meaningful results [3]. In addition, every 1-gram frequency in the years 1899 and 1905 was replaced by the average for that 1-gram from ±1 years, as the *Google Books Ngram Corpus* occasionally assigned 1899 or 1905 to books of unknown publication dates [3]. For any 1-gram that appeared in year $t$ but not in $t-1$ and $t+1$ to $t+10$ against what expected from the usual patterns (Figure S1), we set its frequency in $t$ to zero to avoid possible errors from the optical character recognition (OCR) processes.

### 1.3. Identification of scientific and technological words

How do we know whether a given 1-gram belongs to the vocabulary of science and technology? One simple way is to check whether it matches any word in a published science dictionary. We created a list of scientific and technological words (1-grams) from an online science dictionary "*AccessScience*" [4]. Because the list itself may be biased toward words in common use today, we added words from other various sources including those used in the past as well (Table S1): we extracted words from patent grant texts in the *United States Patent and Trademark Office* data provided by Google [5] and from article titles in a number of scientific journals. Among those words, only nouns were selected. A word was considered to be a noun, if it was used as a noun in more than 90% of its total usage in the year 2000 (e.g., 5,377,995 / 7,285,922 = 73.8% usage as a noun for "work" in section '*Google Books Ngram Corpus*'). We filtered out words with the year of birth < 1800 to make them consistent with section 'Preprocessing'. Then, we arranged the remaining words in descending order of usage within their respective sources. For most cases, the words of high usage within the sources were likely to be scientific and technological words. By manual inspection of randomly-sampled words (≥ 10% coverage for journals, ≥ 1% coverage for patents) along the descending order of usage level within each source, we selected all words of the usage level having at least an 80% chance of being scientific and technological words which are not used in too broad a context. If this cutoff covered all words occurring in that source, then we excluded words used only once in the source. In total, we obtained 7,855 scientific and technological words from the dictionary, patents, and journals.

### 1.4. Connection between word usage and events in society

One may ask about how word usage is related to the empirical events in society. We here present several examples in response to such questions. The original study of *Google Books Ngram Corpus* [3] reported that the boost of a word in its frequency can reflect the increasing impact of the relevant event on society. For example, peaks in "influenza" correspond to the dates of known pandemics [3]. Additionally, various studies in sociolinguistics have paid attention to connections between, e.g., social structures and word usage [6]–[7], urbanized population and word usage [8], and events in society and coherent changes in word usage [9]. Those studies seem to support our assumption that the frequency of a scientific word is indicative of the actual impact of the scientific concept on society.

## 2. Data analysis

### 2.1. Determination of $f_c$, FPT, lifetime, and peak

The cutoff frequency $f_c$ defines the threshold above which a 1-gram can be roughly considered to be common in society. A proper choice of $f_c$ is important as the quantification of first passage time and lifetime (see below) depends on it. We chose $f_c = 10^{-7}$ since 1-grams with frequency $> 10^{-7}$ are easily found in published dictionaries [3]. In 2000, there were 79,691 word stems (corresponding to ~200,000 1-grams) with frequency $> 10^{-7}$ (Figure S2). Our main results presented in this work, however, do not qualitatively change as long as $10^{-8} \leq f_c \leq 2\times10^{-7}$. For a given 1-gram, first passage time (FPT) is defined as years to cross $f_c$ in frequency since the birth of the 1-gram, lifetime is defined as years between the first and the last year the frequency was above $f_c$, and peak is defined as the highest frequency of the 1-gram over time. Specifically, we define lifetimes to 1-grams, which are under the frequency $f_c$ for at least 10 years until the year 2008, since they are rarely expected to bounce back (Figure S1). Figure S3 illustrates the definitions of FPT, lifetime, and peak.

### 2.2. Characterization of different 1-gram types

Most 1-grams could be classified into the following three types. Type-I has 1-grams with well-defined finite lifetimes (section 'Determination of $f_c$, FPT, lifetime, and peak'). Type-II shows a lifetime to a distinctively long extent beyond the time frame, so the exact lifetime cannot presently be defined. Type-III, unlike types-I and -II, never had a frequency higher than $f_c$. One may claim that the distinction between type-I and type-II was merely based on the limited period of observation allowed in our current dataset. Although the distinction was made in a rather heuristic way, we did observe a more fundamental difference between type-I and type-II. Figure S4 shows the probability density function (PDF) of the frequency for each type of 1-grams in a given year. While the PDFs for type-I and type-II initially overlap, the difference between them grows over time as the PDF of type-II shifts to higher frequency ranges. The growing difference can be quantified by tracking the average and median frequencies of each type over the years, as shown in Figure S5. While the average and median frequencies of type-I stay almost steady, the same statistics of type-II keep increasing. The results indicate an intrinsic difference between types-I and -II, manifested in their frequency growth patterns.

## 2.3. Predictability

Type-II includes scientific words prevailing in society longer than the other types. Thus, by identifying type-II scientific words at a relatively early stage, we can predict which words will be promising in the future. As demonstrated in Figure S5, the frequency of a type-I word tends to stay at a low level, while that of a type-II word continuously grows. This fact implies that if we identify the scientific words whose frequency exceeds a sufficiently high level, many of them will be type-II. Figure S6 indeed shows that the higher the level of frequency exceeded, the more likely the word belongs to type-II. It also shows that the probability of being type-II varies slightly across the years when the words passed a particular level of frequency. This raises the question of which years are appropriate to choose to estimate the precision of type-II identification. The period of the years should be long enough for a reliable statistical analysis and the years should be old enough for a clear distinction between type-I and type-II in 2008. We selected the period of years between 1800 and 1919, which leaves 89 years until the end year of our dataset, and this 89-year period is longer or comparable to the typical lifespan of a human being.

For the period 1800–1919, the relationship between the level of frequency exceeded and the probability of being type-II in 2008 is presented in Figure 1b. Accordingly, we made a list of scientific words predicted to be future type-II based on the level of frequency passed in the years between 2000 and 2008 (Tables S2–S5). All entries were classified into respective categories, and we filtered out the words used in too broad a context, not necessarily in a scientific context.

### 2.3.1. Significance test

To test the statistical significance of the relation between the level of frequency passed and the probability of being type-II, we performed a two-sided Z-test under the null hypothesis that there is no association between the frequency level and the probability of type-II, resulting in their correlation merely by chance. For this analysis, we calculated expected values and standard deviations from the null distributions. Among $N$ scientific words (1-grams) in total, let $q$ be the fraction of words over a certain frequency level and $r$ be the fraction of type-II. The expected number of type-II over the frequency level is $Nqr$ and the variance is $Nqr(1 − r)$. The central limit theorem ensured that this null distribution converged well to the Gaussian distribution, giving a Z-score as well as a P-value (Table S6).

## 2.3.2. Internet webpage volume

To test the validity of our type-II prediction results against an up-to-date independent dataset, we used the Google web search engine that showed the Internet webpage volumes updated annually between 2008 and 2013 for the words of our search queries (accessed in February and March 2014). Because Google provides search results using a stemming algorithm, we submitted the singular forms of the words instead of the word stems themselves. Because Google does not permit automatic search queries by web robots, we manually submitted (i) the type-II-predicted scientific words in Tables S2–S4, and (ii) their counterparts, randomly-selected from the scientific words that first reached any frequency $\leq 2\times10^{-6}$ between 2000 and 2008. For the normalization in Figure 1c, we used 100 random words from (ii). For the control group against (i) in Figure 1d, we used 100 random words from (ii) not overlapping with (i). In Figure 1d, the comparison between the search queries for (i) and for the control group shows that the prediction results also work for the webpage volumes since 2008, although the prediction itself is based on the 1-gram data between 2000 and 2008.

## 2.4. Rescaling of FPT and lifetime

Figure S7 gives a detailed visualization of the results in Figure 2a: for each set of type-I 1-grams born in the same year, the complementary cumulative distribution functions (CCDFs) of FPT, lifetime, and their rescaled values (FPT and lifetime divided by their respective averages from the same year of birth) are plotted. Overall FPT and lifetime were getting shorter over the past years, as the CCDFs of FPT and lifetime were getting steeper as the years passed. Rescaled values lead to a collapsing of their CCDFs from different years into an approximately single curve, indicating nearly equivalent patterns are followed across years for FPT and lifetime. The rescaling doesn't only work for all type-I 1-grams, but also separately for the type-I scientific words among them (Figure S8).

## 2.5. Relations between FPT, lifetime, and peak

This section discusses the unique features of scientific words manifested in the relations between FPT, lifetime, and peak. For FPT and lifetime in this section, we use their rescaled values (section 'Rescaling of FPT and lifetime') unless specified.

### 2.5.1. Adjusted density plot

To find the correlation between two quantities, $x$ and $y$ in the linear scale, we first take a small window of size $b_x \times b_y$, place the lower left corner of the window at the starting (smallest) points of $x$ and $y$ ($x_{min}$, $y_{min}$), measure the density of data points inside the window, and assign the value to the lower left corner. We repeat the same procedure after shifting the position of the window by $b_x/k_x$ along the $x$-axis or $b_y/k_y$ along the $y$-axis until the entire $xy$-plane is spanned ($k_x$ and $k_y$ are constants). If one axis (say $x$) is in the logarithmic scale, the density at each position is calculated in a similar way except that the window is shifted in the $x$-direction by multiplying $(i_x)^{1/k_x}$ to the $x$-coordinate and the window length along the $x$-axis increases by the same factor. Finally, we normalize every density at each $x$ relative to the maximum across the $y$-axis. We call this density "adjusted density", which is suited for clarifying the dependence of $y$ on $x$ when plotted on the $xy$ plane.

### 2.5.2. FPT and lifetime

Figure S9 (same as Figure 2b and c) shows the density plot between FPT and lifetime, for scientific words (left) and an entire set of 1-grams (right) in type-I. For scientific words, FPT and lifetime are negatively correlated, with a transition at FPT~1.2 giving rise to a sudden appearance of lifetime~2.0 (Pearson's Chi-squared test, $P = 4.3 \times 10^{-47}$). For an entire set of 1-grams, there is no such transition.

### 2.5.3. Peak and lifetime

Figure S10 shows the density plot between peak and lifetime for scientific words (left) and an entire set of 1-grams (right) in type-I. At small values of peak for scientific words, lifetimes are mostly short. As peak increases, a sudden leap from short to long lifetime is observed at peak ~ $5 \times 10^{-7}$. This transition barely occurs for an entire set of 1-grams, at much larger peak (11.3 times larger) than for scientific words.

### 2.5.4. FPT and peak

Figure S11 shows the density plot between FPT and peak for scientific words (left) and an entire set of 1-grams (right) in type-I. FPT and peak have negative correlation.

### 2.5.5. Significance test

To test the statistical significance of sudden leap into ~2.0 in lifetime at FPT~1.2 for type-I scientific words, we constructed a $2 \times 2$ contingency table displaying the numbers of the words at FPT $\geq$ 1.2 and < 1.2, and lifetime $\geq$ 2.0 and < 2.0. Then, we computed the Pearson's Chi-square value and a $P$-value based on a Chi-square distribution with 1 degree of freedom, with a null hypothesis that there is no association between FPT and lifetime.

## 3. Model description

To build a mechanistic model to account for our observation, we considered the three key factors in the spread of science and technology – preferential adoption, homophily, and fitness, as described in the main text. In this section, we explain further details of how the model accommodates these factors. The model consists of $N$ agents where individual agents represent various forms of social units to invent and adopt items. The items are transmitted from agent to agent. We assume that the adopted ranges of such items are projected into the actual usage levels of the corresponding words in the 1-gram dataset [3].

### 3.1. Homophily

Each agent is assigned $\varepsilon$, which characterizes the level of involvement in specialized areas. In general, $\varepsilon$ can be a vector with real-number components. For the simplicity of our model, here $\varepsilon$ is a scalar binary number: $\varepsilon = 1$ if the agent belongs to the scientific community, otherwise, $\varepsilon = 0$. In other words, agents such as scientists, engineers, scientific journalists, research institutes, and scientific publishers can take $\varepsilon = 1$, and we call them simply 'scientists' in our model. Scientists occupy only a small fraction of the whole system, with a certain chance of being a scientist (equal to $\rho$) given to each agent at the beginning of the simulation. Once $\varepsilon$ has been determined to be either $\varepsilon = 1$ or 0 for each agent, it never changes during the simulation. To consider the effect of homophily, we introduce a weight function for every pair of agents, $w(|\varepsilon_i - \varepsilon_j|)$, which captures how influential agents $i$ and $j$ in the pair are to each other in the spread of innovation. $w(|\varepsilon_i - \varepsilon_j|)$ should be a decreasing function of $|\varepsilon_i - \varepsilon_j|$ and we chose the form $w(|\varepsilon_i - \varepsilon_j|) = \exp[-(\varepsilon_i - \varepsilon_j)^2]$.

### 3.2. Preferential adoption and homophily

When agent $i$ adopts another $j$'s item $q$, preferential adoption and homophily work as the following function, $p(q, i) \times p(q, j)$, where

$$p(q,m) = \sqrt{\frac{\sum_r \delta(q,r) w(|\varepsilon_m - \varepsilon_r|)}{\sum_r w(|\varepsilon_m - \varepsilon_r|)}}.$$

Here, $\sum_r$ denotes the sum over all agents in the system and $\delta(q, r)=1$ if agent $r$ holds item $q$, otherwise, $\delta(q, r)=0$. $w(|\varepsilon_m - \varepsilon_r|)$ comes from section 'Homophily'. A square root appears in $p(q, m)$ because it makes $p(q, i) \times p(q, j)$ linearly proportional to the population having item $q$ in the case that $\varepsilon$'s are identical for all agents.

### 3.3. Network for information spread

Adoption of new items takes place through direct information spread between agents. For the simulation results presented in this study, the global network topology of such information channels connecting different agents was set following the Erdős–Rényi model [10]. Specifically, we used a $G(N, p_{ER})$ model where each agent is randomly linked to another with probability $p_{ER}$ [11]. To avoid generating isolated agents, we took $p_{ER} > \ln(N)/N$.

We also considered another network model, the static model of scale-free networks [12], which is known to produce a fat-tailed, power-law degree distribution in contrast to the Erdős–Rényi model. For the degree exponent between 2.0 and 3.0 (other parameters set equal to those of Figure 3a–d), we found that our main results did not much change with the selection of this network topology.

### 3.4. Fitness

To each invented item, we assign fitness $\lambda$, which gives the intrinsic differences between items in their adoption rates.

#### 3.4.1. Gaussian distribution

Provided that fitness $\lambda$ is a sum of numerous uncorrelated properties of an item, one can assume that the fitness distribution follows the Gaussian distribution $\Lambda_g(0.5, \sigma) \sim \exp[(\lambda-0.5)^2/2\sigma^2]$, whose domain is centred at 0.5 and bounded by 0 and 1. In this case, we consider the following function contributing to the probability that a new item $q_j$ with fitness $\lambda_{q_j}$ replaces an old item $q_i$ with fitness $\lambda_{q_i}$ in its adoption:

$$f(\lambda_{q_j} - \lambda_{q_i}) = \frac{1}{2} + \left(\frac{\lambda_{q_j} - \lambda_{q_i}}{2}\right).$$

### 3.4.2. Power-law distribution

Alternatively, one can assume that the fitness distribution follows a fat-tailed distribution such as a power-law, $\Lambda_p(\gamma, x_{min}) \sim (x/x_{min})^{-\gamma}$, whose domain is bounded by 1 and 11. In this case, we consider the following function contributing to the probability that a new item $q_j$ with fitness $\lambda_{q_j}$ replaces an old item $q_i$ with fitness $\lambda_{q_i}$ in its adoption:

$$f(\lambda_{q_j} - \lambda_{q_i}) = \frac{1}{2} + \left(\frac{\lambda_{q_j} - \lambda_{q_i}}{10}\right)^\beta \quad \text{if} \quad \lambda_{q_j} \geq \lambda_{q_i},$$

$$f(\lambda_{q_j} - \lambda_{q_i}) = \frac{1}{2} - \left(\frac{\lambda_{q_i} - \lambda_{q_j}}{10}\right)^\beta \quad \text{if} \quad \lambda_{q_j} < \lambda_{q_i}.$$

### 3.5. Update rule

In our model, every agent has $L$ distinct items at every instant. At every time step, a new item is introduced by randomly-selected agent $i$ with probability $\alpha$, and is assigned the category simply by following agent $i$'s specialty $\varepsilon_i$ (section 'Homophily'). The new item randomly replaces one of the agent $i$'s old items in the same category as the new one. If there is no such item in the same category, any old item of agent $i$ is randomly chosen and replaced. The new item has fitness with the probability distribution mentioned in section 'Fitness'.

Next, we randomly select a pair of agents $j$ and $k$ in direct contact through pre-assigned information channels (section 'Network for information spread') and their items $q_j$ and $q_k$ belonging to the same category. If agents $j$ and $k$ have no items in the same category, any pair of their items is selected. Then, agent $j$ adopts item $q_k$ by replacing item $q_j$ with the following probability, provided that agent $j$ has never adopted item $q_k$ before:

$$P(q_j, q_k, j, k) = f(\lambda_{q_k} - \lambda_{q_j}) \times p(q_k, j) \times p(q_k, k),$$

where $\lambda_q$ is item $q$'s fitness, and each function is described in the previous sections. If $P(q_j, q_k, j, k)$ is smaller than 0 (larger than 1), we consider it to be 0 (to be 1). At every $N \times L$ repetitions of the above steps, the frequencies of all items in the system are recorded. The frequency of an item is defined as the ratio of the item's copy number to the total counts of items (= $N \times L$) in the system. Here, we use such $N \times L$ repetitions of the steps as the arbitrary unit of time to measure the FPT and lifetime of items.

### 3.6. Initialization

After the system is set up with given parameters, we start with $N$ agents having no items. We run the simulation as described in section 'Update rule', except that a transmitted (newly generated) item is appended to the receiving (producing) agent's item list if the list contains fewer than $L$ distinct items. If the receiving (producing) agent already has $L$ distinct items, then one of them is replaced with the transmitted (newly generated) item according to the rules in section 'Update rule'. The initialization process is complete once every agent has $L$ distinct items, and the simulation time starts at that moment.

## 3.7. Ergodicity

In section 'Data analysis', all statistics for the 1-gram data were obtained from the long time series data. For the model analysis here, we use the ensemble results assembled from multiple simulations rather than use the results from a single long simulation, to save simulation times. Simulations for each ensemble were performed under the same model parameters but can have different initial conditions and network connectivity due to the randomness in the initialization process. One may question the validity of using such ensemble results instead of results from a sheer long-time simulation. We claim that our model is ergodic enough so that both ensemble and long-time results give almost equivalent patterns. Two Erdős–Rényi networks with equal $p_{ER}$ do not have much statistical difference in their structural properties when the network size is large enough [11], so their dynamical properties would not be much different either. Moreover, most items cannot survive over the frequency $f_c$ in the system for longer than 50000 steps, and within 100000 steps all items of the system fall below $f_c$ and are effectively replaced by the new, not leaving much trace of the past. Therefore, a long simulation of our model would be nearly equivalent to an ensemble of different simulations.

# 4. Model results

The simulation of our model shows, whether for the scientific category or not, the existence of type-II-like items having distinctively longer lifetimes than the others (Figures S12–S14; see also section 'Distinct dynamics of type-I and type-II in their adoption'). They appear even if all agents and items are assigned the same $\varepsilon$ and the same $\lambda$, respectively, indicating that preferential adoption is sufficient for the existence of type-II (Figure S14 for the same $\lambda$ case). However, homophily and fitness effects are also important to explain the observed patterns in scientific words, as discussed below.

## 4.1. Relation between FPT and lifetime

In Figures S12–S14, we show density plots between FPT and lifetime for different forms of fitness distributions, which supplement the results in Figure 3a and b. We checked the cases of the power-law (Figure S12), Gaussian (Figure S13), and Dirac delta (i.e., identical fitness for all items; Figure S14) fitness distributions. In the figures, each top panel is a zoomed-in view of the bottom panel, a region below the boundary made by the constant level of the sum of FPT and lifetime. This boundary imitates the limits of the total time frame of the real 1-gram data, defining a type-I case that indeed captures the FPT–lifetime relationship of type-I observed empirically in Figure 2b and c.

For all three different fitness distributions, we could identify the range of parameters in which (i) type-I and type-II items are clearly distinguishable and (ii) type-I scientific items exhibit the sudden transition of lifetime across FPT. If we don't consider preferential adoption, homophily, and fitness for our model, then the functional form of $P(q_j, q_k, j, k)$ in section 'Update rule' is changed into $P(q_i, q_j, i, j) = \theta$, where $\theta$ is an arbitrary constant. In this case, the feature (i) is observed at a very narrow range of $\theta$, e.g., only at $\theta \sim 0.01$ in the same condition as Fig. 3a–d. If we now consider preferential adoption, the feature (i) appears easily without such parameter fine-tuning. However, preferential adoption alone is not enough for the feature (ii), as the feature (ii) does not appear if all agents have identical $\varepsilon$'s. Therefore, for the features (i) and (ii), preferential adoption and homophily are both important. It is noteworthy that features (i) and (ii) can be produced, even with the Dirac delta distribution of the fitness. Nonetheless, fitness is also important in our model, as the negative correlation between FPT and lifetime in the regime of (rescaled) long lifetime ≥2.0 in Figure 2b after the transition, herein called feature (iii), cannot be reproduced under the Dirac delta distribution of fitness (Figure S14). Therefore, the three fundamental components in the model – preferential adoption, homophily, and fitness – are important to explain the observed patterns in scientific evolution.

For Figure 3a–d, we used the power-law fitness distribution with parameters $\gamma = 2.0$, $\beta = 1/4$, $N = 4{,}096$, $L = 10$, $p_{ER} = 0.1024$, $\rho = 0.2$, $\alpha = 0.0001$, $f_c = 0.00025$. We found that such model outcomes do not qualitatively change, at least in the following parameter ranges: (given $N = 4{,}096$, $L = 10$, and $\beta = 1/4$) $0.1 \leq \rho \leq 0.3$, $0.00001 \leq \alpha \leq 0.001$, $0.0002 < f_c < 0.0003$, $0.0256 \leq p_{ER} < 0.2048$ [for (ii); power-law fitness distribution] or $0.0064 \leq p_{ER} < 0.01$ [for (ii); Gaussian and Dirac delta fitness distributions], $2 \leq \gamma \leq 2.4$ [for (iii); power-law fitness distribution] or $0.07 \leq \sigma \leq 0.13$ [for (iii); Gaussian fitness distribution].

### 4.1.1. Significance test

To test the statistical significance of the sudden transition of lifetime across FPT for type-I scientific items, we conducted an analysis similar to section 'Significance test' in 'Data analysis': in Figure 3a, an abruptly long lifetime ~ 2,000 appears at FPT ~ 5,000. We constructed a $2 \times 2$ contingency table displaying the numbers of the words at FPT $\geq 5{,}000$ and $< 5{,}000$, and lifetime $\geq 2{,}000$ and $< 2{,}000$. Then, we computed the Pearson's Chi-square value and a $P$-value based on a Chi-square distribution with 1 degree of freedom, with a null hypothesis that there is no association between FPT and lifetime ($P = 5.4 \times 10^{-22}$).

### 4.2. Distinct dynamics of type-I and type-II in their adoption

The right panels of Figures S12–S14 show clear gaps between short and long lifetimes of non-scientific items, giving a straightforward way to split type-I and type-II at lifetimes ~ 12,000, 8,400, and 8,000 for Figures S12–S14, respectively. We assume that these values of lifetime to split type-I and type-II are approximately equal for both non-scientific and scientific items. Based on this assumption, we split type-I from scientific items along the boundaries defined in the legends of Figures S12–S14 (see also section 'Relation between FPT and lifetime'). One may question the validity of this classification scheme of type-I and type-II for scientific items, as the left bottom panels of Figures S12–S14 show less clear gaps between type-I and type-II than the right bottom panels. Nonetheless, we were able to demonstrate that type-I and type-II scientific items are qualitatively different in their dynamics.

Figure S15 shows the probability distributions of $\Delta t_f = t_f' - t_f$ for type-I and type-II scientific items, where $t_f'$ ($t_f$) of each item denotes the last time that item's frequency outside (inside) the scientific community fell below $f_c$. $\Delta t_f > 0$ ($< 0$) indicates that the item has been in longer common use outside (inside) than inside (outside) the scientific community. In Figure S15, we observe that type-I and type-II tend to occupy different regimes of $\Delta t_f$: $\Delta t_f < 0$ for type-I and $\Delta t_f > 0$ for type-II. In other words, type-II scientific items tend to survive in the outer society even though they are no longer active within the scientific community. The adoption of type-I scientific items shows the opposite trend, largely driven by the internal dynamics of the scientific community itself. In conclusion, the simulation results demonstrate the fundamental difference between type-I and type-II in their dynamics during adoption, supporting the validity of our classification scheme for type-I and type-II. For this analysis, we excluded the items whose frequency either outside or inside the scientific community never exceeded $f_c$, because of their ill-defined $t_f'$ and $t_f$. These items would not be well found near the boundaries between type-I and type-II, so excluding them would not distract the rigorous examination of the difference between these two types.

### 4.3. Effect of fitness on lifetime and peak

In our model, the spread of an item depends on its fitness as well as social effects. The latter effects do not always favor the spread of a higher-fitness item because they may amplify random fluctuations in the item's spread and strengthen the spread of the item in the majority regardless of its fitness. In this section, we present simulation results showing how critical fitness is in determining the long-term fate of individual items – lifetime and peak.

Figure S16 shows the averages of lifetime and peak steadily increasing over fitness, but also the large variability out of this average trend. In Figure 3c and Figure S17, we use the coefficient of variation (CV) as an indicator of the variability. CV is defined as the ratio of standard deviation to mean. Figure S17 shows that the variability of lifetime and peak increases non-monotonically across fitness, reaching the maximum at the intermediate level of fitness. In other words, the long-term fate of scientific items is less variable at low and high fitness, and actually type-II and type-III have distributions biased to these fitness regimes (type-II for high fitness and type-III for low fitness; Figures S17–S18).

## 5. Late bloomers: effect of fitness on FPT

Common intuition suggests that FPT and fitness should be anti-correlated. On the contrary, the simulation results clearly show the positive correlation between them for types-I and -II scientific items (Figure S19). These counter-intuitive results can be explained by the fact that high fitness helps the science survive long periods of frequency $< f_c$, allowing for long FPT as well as short FPT (Figure S20). In contrast, low-fitness science is difficult to survive unless it initially spreads fast, either having short FPT or falling to type-III (Figure S20). The existence of high fitness, long FPT science reminds us of the concept 'late bloomers'.

We found that such model outcomes do not qualitatively change, at least in the following parameter ranges: (given $N = 4,096$, $L = 10$, and $\beta = 1/4$) $0.1 \leq \rho \leq 0.3$, $0.00001 \leq \alpha \leq 0.001$, $0.0002 < f_c < 0.0003$, $0.0064 \leq p_{ER} < 0.8192$ (power-law fitness distribution) or $0.0064 \leq p_{ER} < 0.01$ (Gaussian fitness distribution), $2 \leq \gamma \leq 2.4$ (power-law fitness distribution) or $0.07 \leq \sigma \leq 0.13$ (Gaussian fitness distribution).

The above findings raise the possibility that scientific words with very long but finite FPT in the *Google Books Ngram Corpus* dataset can be good candidates for late bloomers with high fitness. We listed in Table S7 such late bloomer candidates from type-II scientific words with rescaled FPT $\geq 2.0$. For this, we manually excluded the words involving dating or OCR errors, and non-scientific use.

### 5.1. Significance test

To test the statistical significance of a positive correlation between FPT and fitness, we performed a two-sided Z-test under the null hypothesis that there is no association between FPT and the fraction of scientific items with high fitness >10.5 (Figure 3d; power-law fitness distribution with $\gamma = 2.0$, $\beta = 1/4$, $N = 4{,}096$, $L = 10$, $p_{ER} = 0.1024$, $\rho = 0.2$, $\alpha = 0.0001$, $f_c = 0.00025$). We calculated an expected value and a standard deviation from the null distribution. For types-I and -II scientific items, let $q$ be the fraction of FPT > 10000 and $r$ be the fraction of fitness > 10.5. If there are $N$ items in total, the expected number of items with fitness > 10.5 among those with FPT > 10000 is $Nqr$ and the variance is $Nqr(1-r)$. The central limit theorem ensured that this null distribution converged well to the Gaussian distribution, giving a Z-score as well as a P-value.

# 6. Evolution of other fields

Our model predicts that other innovative fields such as food and art have similar features to science in FPT-lifetime relation, as demonstrated in Figure 4. However, one of the sources from which we collected food-related words [13] contained 43 (out of 236) type-I words overlapping with those analysed for scientific evolution. To avoid any possible artifact in Figure 4a made by such overlapping, we repeated the analysis after excluding the overlapping words and again found a result similar to Figure 4a (Figure S21; Pearson's Chi-squared test, $P = 4.9 \times 10^{-7}$).

### 6.1. Significance test

To test the statistical significance of the sudden leap in lifetime across FPT in Figure 4 and Figure S21, we conducted an analysis similar to section 'Significance test' in 'Relations between FPT, lifetime, and peak'.

# Supplementary References